\documentclass[10pt,aps,preprintnumbers,prd,noshowpacs,nofootinbib,noshowkeys,floatfix,superscriptaddress]{revtex4}
\usepackage[dvips]{graphics,graphicx}
\usepackage[colorlinks=true,linktocpage=true,linkcolor=blue,citecolor=blue]{hyperref}
\usepackage[usenames,dvipsnames]{color}
\usepackage{amsmath, amssymb,oldgerm}
\usepackage{multirow}
\usepackage{longtable}
\usepackage{color}
\usepackage[normalem]{ulem}  
\usepackage{braket}
\usepackage{slashed}
\usepackage[mathscr]{eucal}

\newcommand{\Tr}{\text{Tr}}
\newcommand{\n}{\nonumber}

\newcommand{\F}{\mathcal{F}}
\newcommand{\Pc}{\mathcal{P}}
\newcommand{\V}{\mathcal{V}}
\newcommand{\A}{\mathcal{A}}
\newcommand{\mC}{\mathfrak{C}}
\newcommand{\C}{\mathcal{C}}
\newcommand{\Sc}{\mathcal{S}}
\newcommand{\ms}{\mathfrak{s}}
\newcommand{\f}{\mathfrak{f}}

\newcommand{\beq}{\begin{equation}}
\newcommand{\eeq}{\end{equation}}
 \newcommand{\im}{\text{Im}\, }
 \newcommand{\re}{\text{Re}\, }

\newcommand{\olra}{\overleftrightarrow}
\newcommand{\eqs}{Eqs.~}
\newcommand{\eq}{Eq.~}

\newcommand{\inlangle}{{\phantom{\Big|}}_\text{in}\Big\langle}
\newcommand{\inrangle}{\Big\rangle_\text{in}}
\newcommand{\inlanglesm}{{}_\text{in}\langle}
\newcommand{\inranglesm}{\rangle_\text{in}}
\newcommand{\outlangle}{{\phantom{\Big|}}_\text{out}\Big\langle}
\newcommand{\outrangle}{\Big\rangle_\text{out}}
\newcommand{\outlanglesm}{{}_\text{out}\langle}

\newcommand{\os}{\text{on-shell}}
\newcommand{\osl}{\text{on-shell},l}

\begin{document}

\title{Derivation of the nonlocal collision term in the relativistic Boltzmann equation for massive spin-1/2 particles from quantum field theory}

\author{Nora Weickgenannt}

\affiliation{Institute for Theoretical Physics, Goethe University,
Max-von-Laue-Str.\ 1, D-60438 Frankfurt am Main, Germany}

\author{Enrico Speranza}

\affiliation{Institute for Theoretical Physics, Goethe University,
Max-von-Laue-Str.\ 1, D-60438 Frankfurt am Main, Germany}
\affiliation{Illinois Center for Advanced Studies of the Universe and Department of Physics, University of Illinois at Urbana-Champaign, Urbana, IL 61801, USA}
\affiliation{Helmholtz Research Academy Hesse for FAIR, Campus Riedberg, Max-von-Laue-Str.\ 12, D-60438 Frankfurt am Main, Germany}

\author{Xin-li Sheng}

\affiliation{Key Laboratory of Quark and Lepton Physics (MOE) and Institute of Particle Physics, Central China Normal University, Wuhan, 430079, China}

\author{Qun Wang}

\affiliation{Interdisciplinary Center for Theoretical Study and Department of Modern Physics, University of Science and Technology of China, Hefei, Anhui 230026, China}
\affiliation{Peng Huanwu Center for Fundamental Theory, Hefei, Anhui 230026, China }

\author{Dirk H.\ Rischke}

\affiliation{Institute for Theoretical Physics, Goethe University,
Max-von-Laue-Str.\ 1, D-60438 Frankfurt am Main, Germany}
\affiliation{Helmholtz Research Academy Hesse for FAIR, Campus Riedberg, Max-von-Laue-Str.\ 12, D-60438 Frankfurt am Main, Germany}

\begin{abstract}
We derive the Boltzmann equation and the collision kernel  
{for massive spin-1/2 particles}, using the Wigner-function formalism 
and employing an expansion in powers of $\hbar$. The phase space is enlarged to 
include a variable related to {the} spin degrees of 
freedom. This allows to reduce the transport equations of the independent 
components of the Wigner function to one scalar equation. 
To next-to-leading order in $\hbar$, we find that the collision kernel contains 
both local and nonlocal terms. We show that off-shell contributions cancel in the 
Boltzmann equation. 
Our framework can be used to study spin-polarization phenomena induced by vorticity 
as recently observed in heavy-ion collisions and in condensed-matter systems.
\end{abstract}

\preprint{USTC-ICTS/PCFT-20-13}



\maketitle

\section{Introduction}

Polarization phenomena in nuclear collisions have been {recently} 
the focus of intense research. The large orbital angular momentum in noncentral 
heavy-ion collisions can {(at least partially) be transferred as vorticity} to the 
hot and dense matter created in the collision zone. This, in turn, may align the spins of 
particles along the direction of {the} global orbital angular momentum, 
leading to a nonzero spin polarization~\cite{Liang:2004ph,Voloshin:2004ha,Betz:2007kg,Becattini:2007sr}. 
Such a mechanism is rather similar to the time-honored Barnett effect \cite{Barnett:1935}. 
In a nonrelativistic system, the alignment of spins by rotation implies the alignment of 
magnetic moments and thus polarization is equivalent to magnetization. In a relativistic 
system, however, both particles and antiparticles are present,
and while the spins of particles and antiparticles align in the same direction through 
rotation, the magnetic moments align in the opposite direction, which reduces the 
magnetization. Thus, a system with equal
numbers of particles and antiparticles is polarized, but not magnetized.
This effect is a prime example for the interplay between a macroscopic quantity, the 
rotation, and a microscopic quantity, which is inherently of quantum nature: the spin
of the particles. 

The STAR Collaboration found that Lambda baryons emitted in noncentral heavy-ion 
collisions are indeed emitted with a finite global polarization (i.e., a polarization along the 
direction of the global angular momentum), 
providing evidence of spin polarization generated by vorticity~\cite{STAR:2017ckg}. 
The global polarization predicted by models based on the assumption of local 
thermodynamic equilibrium of spin degrees of freedom 
{turn} out to be in good agreement with the experimental findings~\cite{Becattini:2007sr,Becattini:2013vja,Becattini:2013fla,Becattini:2015ska,Becattini:2016gvu,Karpenko:2016jyx,Pang:2016igs,Xie:2017upb}. 
More recently, the STAR Collaboration measured the projection of the Lambda 
polarization along the beam direction, the so-called longitudinal polarization, as a function 
of the azimuthal angle of the particles~\cite{Adam:2019srw}. Unfortunately, the same 
theoretical models which were able to describe the global-polarization data 
predict an opposite sign with respect to the experimental observations, often dubbed the 
``polarization sign problem''~\cite{Becattini:2017gcx,Becattini:2020ngo}.
A number of attempts~\cite{Florkowski:2019qdp,Florkowski:2019voj,
Zhang:2019xya,Becattini:2019ntv,Xia:2019fjf,Wu:2019eyi,Sun:2018bjl,Liu:2019krs} 
have been made to explain the polarization sign problem, but as of yet no definite 
conclusion has been reached.

The polarization sign problem suggests that spin degrees of freedom have nontrivial 
dynamics, which is not captured by the theoretical models used to accurately describe the
global-polarization data. One possibility is that nonequilibrium effects of spin degrees of 
freedom have to be included in the kinetic and hydrodynamic description of the hot and 
dense matter. A theory of relativistic spin hydrodynamics, first introduced in 
Ref.~\cite{Florkowski:2017ruc,Montenegro:2017rbu} and followed by Refs.\ \cite{Florkowski:2017dyn,Florkowski:2018myy,Becattini:2018duy,Florkowski:2018fap,Montenegro:2017lvf,Hattori:2019lfp}, 
has {also been} recently {derived from} various approaches: 
kinetic theory~\cite{Bhadury:2020puc,Weickgenannt:2020aaf,Shi:2020htn,Speranza:2020ilk,Bhadury:2020cop,Singh:2020rht,Bhadury:2021oat}, 
effective action~\cite{Montenegro:2018bcf,Montenegro:2020paq,Gallegos:2021bzp}, 
entropy-current analysis~\cite{Fukushima:2020ucl,Li:2020eon},
and holographic duality~\cite{Gallegos:2020otk,Garbiso:2020puw}. 

From a microscopic point of view, the dynamics underlying the conversion between 
orbital and spin angular momentum (and vice versa) can be understood in terms of 
particle collisions. Such dynamics was studied in the nonrelativistic case in a seminal 
paper by Hess and Waldmann, where a kinetic theory for a dilute gas of particles 
with spin was formulated~\cite{hess1966kinetic}. One of the main conclusions of this 
work is that, in order to describe polarization phenomena through rotation (e.g.\ the 
Barnett effect), one needs nonlocal particle collisions. The authors were not able to 
provide a first-principle derivation of the nonlocal collision kernel and 
they phenomenologically added terms in their kinetic theory to describe the 
orbital-to-spin angular momentum 
conversion. A nonrelativistic Boltzmann equation with a nonlocal collision term 
was then discussed by Hess in Ref.\ \cite{hess1967verallgemeinerte}.
Detailed derivations of nonlocal collision terms for a nonrelativistic system of spinless 
particles can be found, e.g., in Refs.~\cite{Morawetz:2017blm,Morawetz:2018shf,Morawetz:2000hr}. 
In the relativistic case, {a microscopic mechanism based on nonlocal 
scatterings 
between wave packets was proposed in Ref.~\cite{Zhang:2019xya}} to explain the 
generation of the spin-vorticity coupling in heavy-ion collisions. However, to the best of 
our knowledge, {a systematic derivation of a 
nonlocal collision kernel in the relativistic Boltzmann equation for particles with spin 
based on quantum field theory} has only been performed very recently in our previous 
work~\cite{Weickgenannt:2020aaf} 
[for related efforts, see also Refs.~\cite{Yang:2020hri,Wang:2020pej}]. 

In Ref.~\cite{Weickgenannt:2020aaf} we presented a Boltzmann equation using the 
Wigner-function formalism, which includes the nonlocality of the scattering process 
between particles and established its connection with spin-hydrodynamics.
In this paper, we now give the details of the derivation.
The Wigner-function formalism provides a first-principle formulation of kinetic theory 
and also {turns} out to be a very powerful tool for the description of 
anomalous transport in the quark-gluon plasma (QGP) created in heavy-ion collisions
[see e.g.\ Refs.~\cite{Son:2012zy,Hidaka:2016yjf,Hidaka:2017auj,Huang:2018wdl,Gao:2018wmr,Yang:2018lew,Gao:2018jsi,Carignano:2018gqt}].
Our derivation is based on a semiclassical expansion of the Wigner function, i.e., an 
expansion in the Planck constant $\hbar$, where spin effects are considered to be of at 
least first order in $\hbar$. 
As it will become clear in the following, an expansion in the Planck constant is also 
effectively an expansion in gradients. Thus, vorticity, which is a quantity of first order in
gradients, is of the same order as spin polarization. The latter can therefore be generated 
from the former through nonlocal scattering processes.

The structure of the paper is the following. In Sec.\ \ref{secqu}, we derive the quantum 
transport equations from the Wigner-function formalism. In order to have a more compact 
transport equation for the components of the Wigner function, in Sec.\ \ref{secspph} we 
enlarge the phase space by introducing a variable related to spin. 
In Secs.\ \ref{secloc} and \ref{secnon} we explicitly derive the local and nonlocal parts of 
the collision term, respectively. Such calculations are based on the method discussed in 
Ref.~\cite{DeGroot:1980dk}. Finally, conclusions are given in Sec.\ \ref{secconc}. 
We use the following notation and conventions: $a\cdot b=a^\mu b_\mu$,
$a_{[\mu}b_{\nu]}\equiv a_\mu b_\nu-a_\nu b_\mu$, 
$g_{\mu \nu} = \mathrm{diag}(+,-,-,-)$,
$\epsilon^{0123} = - \epsilon_{0123} = 1$, and repeated indices are summed over.

\section{Quantum transport equations}
\label{secqu}

We start from the Wigner function for spin-1/2 particles defined as
\cite{DeGroot:1980dk,Heinz:1983nx,Vasak:1987um},
\begin{equation}
  W_{\alpha\beta}(x,p)=\int \frac{d^4y}{(2\pi\hbar)^4} e^{-\frac i\hbar p\cdot y}
  \left\langle :\bar{\psi}_\beta\left(x_1\right)\psi_\alpha\left(x_2\right):\right\rangle \; , \label{wigdef}
\end{equation}
with $x_{1,2}=x \pm y/2$ and $\psi (x),\, \bar{\psi}(x)$ being Dirac spinor fields. {Here $\langle:\, :\rangle$ denotes the normal-ordered ensemble average.}
In our previous work~\cite{Weickgenannt:2019dks} [see also related work in
Refs.~\cite{Fang:2016vpj,Florkowski:2018ahw,Gao:2019znl,Hattori:2019ahi,Wang:2019moi,Liu:2020flb,Manuel:2021oah}] 
we derived general solutions of the equations of motion for the Wigner function in the
free-streaming limit. Here we extend this idea by including collisions and thus account for 
the effect of interactions. The Lagrangian for Dirac fields used in this paper is of the form
\begin{equation}
\mathcal{L}_D=\bar{\psi}\left(\frac{i\hbar}{2}\gamma\cdot\olra{\partial}-m\right)\psi
+\mathcal{L}_I\;, 
\end{equation}
with $\olra{\partial}\equiv \overrightarrow{\partial}-\overleftarrow{\partial}$ and  
$\mathcal{L}_I$ being a general interaction Lagrangian. {We remark that, if $\mathcal{L}_I$ contains gauge-field interactions, \eq\eqref{wigdef} has to be modified to include a gauge link in order to ensure gauge invariance, see, e.g., Ref.~\cite{Vasak:1987um}.}
We obtain the following equation of motion,
\begin{align}
 \left(i\hbar\gamma\cdot \partial-m\right)\psi(x)&=\hbar\rho(x)\; , \label{Dirac}
\end{align}
where $\rho= - (1/ \hbar) \partial \mathcal{L}_I /\partial \bar{\psi} $. 
From Eq.~\eqref{Dirac} one derives 
the transport equation for the Wigner function \cite{DeGroot:1980dk},
\begin{equation}
 \left[ \gamma \cdot \left( p+i \,\frac{\hbar}{2} \partial \right) -m\right] W_{\alpha\beta}
 =\hbar\, \mathcal{C}_{\alpha\beta}\; , \label{Wignerkin}
\end{equation}
where
\begin{equation}
 \C_{\alpha\beta}\equiv \int \frac{d^4y}{(2\pi\hbar)^4} e^{-\frac i\hbar p\cdot y}
 \left\langle :\bar{\psi}_\beta(x_1)\rho_\alpha(x_2):\right\rangle \;.
\end{equation}
By acting  $\gamma\cdot( p+i \,\frac{\hbar}{2} \partial )+m$ onto Eq.~\eqref{Wignerkin} 
and taking the real part, we obtain a modified on-shell condition for the Wigner function
\begin{eqnarray}
 \left(p^2-m^2-\frac{\hbar^2}{4}\partial^2\right)W_{\alpha\beta}(x,p)
 &=& \hbar\,\delta M_{\alpha\beta}(x,p) \; ,
 \label{onshellwiggg}
\end{eqnarray}
with
{
\begin{eqnarray}
 \delta M_{\alpha\beta}&\equiv&\frac{1}{2}\int \frac{d^4y}{(2\pi\hbar)^4}\, 
 e^{-\frac{i}{\hbar}p\cdot y}
 \left\langle :\left[\bar{\rho}(x_1)(i\hbar\gamma\cdot\overleftarrow{\partial} + m)\right]_\beta 
 \psi_\alpha(x_2) 
+\bar{\psi}_\beta\left(x_1\right)\left[ (-i\hbar\gamma\cdot\partial+m) 
\vphantom{\overleftarrow{\partial}}
\rho\left(x_2\right)\right]_\alpha :\right\rangle\; .
 \label{deltaM}
\end{eqnarray}}
On the other hand, from the imaginary part, we find a Boltzmann-like equation for the 
Wigner function,
\begin{equation}
p\cdot\partial\, W_{\alpha\beta}(x,p)=C_{\alpha\beta}(x,p) \;, \label{wigboltz9}
\end{equation}
with{
\begin{eqnarray}
\label{cwww}
C_{\alpha\beta}&=&\frac{i}{2} \int \frac{d^4y}{(2\pi\hbar)^4} e^{-\frac i\hbar p\cdot y}
\left\langle :\left[\bar{\rho}\left(x_1\right)(-i\hbar\gamma\cdot\overleftarrow{\partial}
  +m)\right]_\beta \psi_\alpha\left(x_2\right)
-\bar{\psi}_\beta\left(x_1\right)\left[(i\hbar\gamma\cdot\partial+m)
\vphantom{\overleftarrow{\partial}}
\rho\left(x_2\right)\right]_\alpha :\right\rangle\; .
\end{eqnarray}}
We will restrict the following considerations to the positive-energy part of the
Wigner function. The extension to negative energies is straightforward. 
Thus, in what follows all mass-shell delta functions are implicitly accompanied by a 
$\theta(p_0)$, which we do not explicitly denote for the sake of simplicity. 

In order to reveal the dependence on the Wigner function on the right-hand side of 
Eq.~\eqref{cwww}, it is convenient to calculate the ensemble average by performing 
the trace over the noninteracting initial $n$-particle states defined 
as~\cite{DeGroot:1980dk} 
\begin{equation}
\label{statein}
|p_1,\ldots,p_n;r_1,\ldots, r_n\rangle_\text{in}\equiv a^\dagger_{\text{in},r_1}(p_1)\cdots 
a^\dagger_{\text{in},r_n}(p_n)|0\rangle \; ,
\end{equation}
where $p_i$ and $r_i$, $i=1, \ldots, n$, denote the particle momentum and spin 
projection, respectively, and 
$a^\dagger_{\text{in},r_i}(p_i)$ is the creation operator for that particle. Since we are 
interested in a kinetic description, we neglect initial correlations. This corresponds to the 
molecular-chaos assumption. Furthermore, we restrict ourselves to two-particle states, 
i.e., we only consider binary collisions. 
Hence, Eq.\ \eqref{cwww} can be written in the form \cite{DeGroot:1980dk} 
(see App.~\ref{dercollapp} for details)
\begin{align}
C_{\alpha\beta}={}& \frac{1}{2(4\pi\hbar m^2)^2}\sum_{r_1,r_2,s_1,s_2}
\int d^4 x_{1}d^4 x_{2}d^4 p_{1}d^4 p_{2}d^4 q_{1}d^4 q_{2} \nonumber\\ &\times 
{\phantom{\Big|}}_\text{in}
\Big\langle{p_1-\frac{q_1}{2},p_2-\frac{q_2}{2};r_1,r_2 \Big| \Phi_{\alpha\beta}(p) 
\Big| p_1+\frac{q_1}{2}, p_2+\frac{q_2}{2};s_1,s_2}\Big\rangle_\text{in}\n\\
&\times\prod_{j=1}^2 \exp\left(\frac{i}{\hbar}q_j\cdot x_j\right)
  \bar{u}_{s_j}\left(p_j+\frac{q_j}{2}\right)
  W_\text{in}(x+x_j, p_j)u_{r_j}\left(p_j-\frac{q_j}{2}\right)\; ,    \label{Wcollttt}
\end{align}
where the operator $\Phi$ is given by
\begin{align}
\Phi_{\alpha\beta}(p)& \equiv\frac{i}{2} \int\frac{d^4y}{(2\pi\hbar)^4} 
e^{-\frac i\hbar p\cdot y} 
:\bigg\{\left[P_\mu, \bar{\rho}\left(\frac y2\right) \gamma^\mu\right]_\beta 
\psi_\alpha\left(-\frac y2\right)
+m\bar{\rho}_\beta \left(\frac y2\right) \psi_\alpha\left(-\frac y2\right)\n\\
&\hspace*{3.2cm} -\bar{\psi}_\beta\left(\frac y2\right)
\left[\gamma^\mu\rho\left(-\frac y2\right),P_\mu\right]_\alpha
-m\bar{\psi}_\beta\left(\frac y2\right)\rho_\alpha\left(-\frac y2\right)\bigg\}:\;,
\label{phigen}
\end{align}
where $P^\mu$ is the total 4-momentum operator. {We also introduced the variable $q_j$ which is the conjugate to $x_j$ in the Wigner transformation, but not related to the particle momenta $p_i$. }
We notice that the Boltzmann-like equation \eqref{wigboltz9} with the collision kernel 
\eqref{Wcollttt}  is not a closed equation for the interacting Wigner function $W$, as 
$C_{\alpha\beta}$ is a functional of the initial 
Wigner function $W_\text{in}$. However, for a dilute system, we further approximate
\begin{equation}
\label{ass1}
W=W_\text{in}+ \ldots \; ,
\end{equation} 
where the ellipsis corresponds to corrections of higher order in density, 
which we neglect \cite{DeGroot:1980dk}. 
We will invert this relation and replace $W_\text{in}$ in the collision term by $W$.
Furthermore, we see that the collision term in Eq.\ \eqref{Wcollttt} takes into account the 
nonlocality of the collision process, as the Wigner functions depend on $x+x_j$. 
If the Wigner function varies slowly in space and time on the microscopic scale
corresponding to the interaction range, we can Taylor-expand $W(x+x_j, p_j)$ 
around $x$ and keep only terms up to first order in gradients 
(equivalent to first order in $\hbar$) \cite{DeGroot:1980dk}, i.e.,
\begin{equation}
\label{ass2}
W(x+x_j, p_j)=W(x, p_j)+x_j \cdot \partial W(x,p_j) \;.
\end{equation}
Substituting Eqs.\ \eqref{ass1} and \eqref{ass2} into Eq.\ \eqref{Wcollttt}, it follows that 
\begin{align}
C_{\alpha\beta} ={}& \frac{(2\pi\hbar)^6}{(2 m)^4}\sum_{r_1,r_2,s_1,s_2}
\int d^4 p_{1}d^4 p_{2} d^4 q_{1}d^4 q_{2} \, {\phantom{\Big|}}_\text{in}
\Big\langle{p_1-\frac{q_1}{2},p_2-\frac{q_2}{2};r_1,r_2\Big|\Phi_{\alpha\beta}(p)
\Big|p_1+\frac{q_1}{2}, p_2+\frac{q_2}{2};s_1,s_2}\Big\rangle_\text{in}\n\\
  &\times\prod_{j=1}^2 \bar{u}_{s_j}\left(p_j+\frac{q_j}{2}\right) 
  \left\{W(x, p_j)\delta^{(4)}(q_j)
  -i\hbar\left[\partial_{q_j}^\mu\delta^{(4)}(q_j)\right]\partial_{\mu}W(x,p_j)\right\}
  u_{r_j}\left(p_j-\frac{q_j}{2}\right)\; , 
  \label{generalcollisionterm}
 \end{align}
where we performed the integration over $d^4x_1$ and $d^4 x_2$.
Equation \eqref{generalcollisionterm} is the collision kernel for the 
Boltzmann equation \eqref{wigboltz9}, 
which we will use as a starting point for the explicit computation of collision effects. 

Following Refs.~\cite{Weickgenannt:2019dks,Gao:2019znl,Hattori:2019ahi,Weickgenannt:2020aaf} 
we employ an expansion in powers of $\hbar$ for the Wigner function, i.e., 
we search for solutions of the form
\begin{equation} \label{Whbarexp}
W=W^{(0)}+\hbar W^{(1)}+\hbar^2 W^{(2)} +\mathcal{O}(\hbar^3)\; .
\end{equation}
We notice that, since gradients are always accompanied by factors of $\hbar$, such 
{}{an} expansion is also a 
gradient expansion. We also stress that the gradient expansion of the nonlocal term 
has to be considered as an $\hbar$ expansion, as Eq.\ \eqref{generalcollisionterm} shows. 

Furthermore, in our treatment, we will consider an expansion around equilibrium,
\begin{equation}
\label{approx1}
W=W_\text{eq}+\delta W \; , 
\end{equation}
where $W_\text{eq}$ is the equilibrium Wigner function and $\delta W$ the deviation 
from equilibrium. In our scheme, we always consider $\delta W$ to be at least of 
first order in an expansion in gradients. As a consequence, if we take into account 
only the lowest-order gradient correction in the nonlocal collision term, we can neglect 
contributions from $\delta W$ in the second term in the second line of 
\eq\eqref{generalcollisionterm}, as they would be of higher order in gradients.  

It is now convenient to decompose the Wigner function in terms of a basis of the 
generators of the Clifford algebra
\begin{equation}
W=\frac14\left(\F+i\gamma^5\Pc+\gamma \cdot \V+\gamma^5\gamma \cdot \A
+\frac12\sigma^{\mu\nu}\Sc_{\mu\nu}\right)\;, \label{dec}
\end{equation}
where $\sigma^{\mu \nu} \equiv \frac{i}{2} [\gamma^\mu, \gamma^\nu]$, 
and substitute it into Eq.~\eqref{Wignerkin} to obtain the equations of motion for 
the coefficient functions. The real part of Eq.~\eqref{Wignerkin} yields
 \begin{subequations}
 \label{real2}
\begin{eqnarray}
p\cdot \V -m\F&=& \hbar D_\F \;,\label{F}\\
\frac{\hbar}{2}\partial\cdot  \A+m\Pc&=&-\hbar D_\Pc\;,\label{P}\\
p^\mu \F-\frac{\hbar}{2}\partial_\nu \Sc^{\nu\mu}-m\V^\mu&=& \hbar D^\mu_\V \;,\label{V}\\
-\frac{\hbar}{2}\partial^\mu \Pc+\frac12\epsilon^{\mu\nu\alpha\beta}p_\nu \Sc_{\alpha\beta}
+m\A^\mu
&=&-\hbar D^\mu_\A\;,\label{A}\\
\frac{\hbar}{2}\partial^{[\mu} \V^{\nu]}-\epsilon^{\mu\nu\alpha\beta}p_\alpha \A_\beta
-m\Sc^{\mu\nu}
&=&\hbar D^{\mu\nu}_\Sc\; ,\label{SSS}
\end{eqnarray}
\end{subequations}
while from the imaginary part we obtain
\begin{subequations}
\label{im2}
\begin{eqnarray}
\hbar\partial \cdot \V&=&2\hbar C_\F\; ,\label{Vkin}\\
p \cdot \A&=&\hbar C_\Pc \;,\label{orth}\\
\frac{\hbar}{2}\partial^\mu \F+p_\nu \Sc^{\nu\mu}&=&\hbar C_\V^\mu\; ,\label{B}\\
p^{\mu}\Pc+\frac{\hbar}{4}\epsilon^{\mu\nu\alpha\beta}\partial_\nu \Sc_{\alpha\beta}
&=& -\hbar C^\mu_\A \;,
\label{Skin}\\
p^{[\mu} \V^{\nu]}+\frac{\hbar}{2}\epsilon^{\mu\nu\alpha\beta}\partial_\alpha \A_\beta
&=&-\hbar C^{\mu\nu}_\Sc\; . \label{Akin}
\end{eqnarray}
\end{subequations}
Here we defined $D_i = \re \Tr\, (\tilde{\Gamma}_i \mathcal{C})$,
$C_i = \im \Tr\, (\tilde{\Gamma}_i \mathcal{C})$, $i = \F,\Pc,\V,\A,\Sc$, 
$\tilde{\Gamma}_\F=1$,
$\tilde{\Gamma}_\Pc =-i \gamma_5$, $\tilde{\Gamma}_\V = \gamma^\mu$,
$\tilde{\Gamma}_\A = \gamma^\mu \gamma^5$,
$\tilde{\Gamma}_\Sc= \sigma^{\mu \nu}$.
Note that each coefficient function will obey a modified on-shell condition and a 
Boltzmann-like equation, 
as can be readily seen from Eqs.\ \eqref{onshellwiggg} and \eqref{wigboltz9}.

We now assume that effects related to spin, and consequently to the polarization, 
are at least of first order in $\hbar$.
This excludes the case of a large initial polarization of the system, i.e., we focus on
situations where a nonzero polarization arises only through
scatterings in the presence of a nonvanishing medium vorticity.
Therefore, since $\A^\mu$ is related to the polarization 
vector~\cite{Weickgenannt:2019dks},
its zeroth-order contribution is assumed vanish, $\A ^{(0)\mu}=0$, and consequently, 
from Eq.~\eqref{SSS},
$\Sc ^{(0)\mu\nu}=0$. Equation~\eqref{P} then implies that $\Pc^{(0)} = 0$.
Thus, at zeroth order all pseudoscalar quantities vanish and, as a consequence, also the
collision terms which carry pseudoscalar quantum numbers must vanish
at zeroth order, $D_\Pc ^{(0)}=C_\Pc^{(0)}=0$. 
Using Eqs.~\eqref{P} and \eqref{orth} this, in turn, implies that
\begin{equation}
	\Pc =\mathcal{O}(\hbar^2)\;, \;\;\; p\cdot \A=\mathcal{O}(\hbar^2)\;.   
	\label{spinorthcoll}
\end{equation}
For the vector part, at zeroth order the only vector at our disposal is $p^\mu$, i.e.,
\begin{equation}
D_\V^\mu= p^\mu\delta V+\mathcal{O}(\hbar)\;,
\end{equation}
with a scalar function $\delta V$. Thus, from Eq.\ \eqref{V} we obtain
\begin{equation}
 \V^\mu=\frac{1}{m}p^\mu \bar{\F}+\mathcal{O}(\hbar^2)\;, \label{newV}
\end{equation}
where we defined $\bar{\F}\equiv \F-\hbar\delta V$.
We can extend this definition to any order in $\hbar$ by setting
\begin{equation}
\label{fbar}
\bar{\F} \equiv \frac{m}{p^2} \, \Tr (p\cdot \gamma \,W) \; .
\end{equation}
We note that the assumption of polarization entering at first order in $\hbar$ implies that  
also the axial-vector $D_\A^{(0)\mu}$ and the antisymmetric tensors
$D_\Sc^{(0)\mu\nu}$ and $C_\Sc^{(0)\mu\nu}$ must vanish. 

We can now write down the modified on-shell conditions for $\bar{\F}$ and $\A^\mu$. 
From \eqs\eqref{F}, \eqref{V}, and \eqref{Akin}, the modified on-shell condition for the 
vector component reads
\begin{equation}
\left(p^2-m^2\right) \V^\mu
= \hbar p^\mu D_\F +\hbar m D_\V^\mu+\mathcal{O}(\hbar^2) \;,
\end{equation}
which, from Eq.\ \eqref{newV}, implies
\begin{equation}
\left(p^2-m^2\right)\bar{\F}= \hbar \, \delta M_F +\mathcal{O}(\hbar^2)
= \hbar m \left(D_\F+\frac{m}{p^2}\,p\cdot D_\V\right)+\mathcal{O}(\hbar^2)  \; , 
\label{Fbaronshell}
\end{equation}
where $\delta M_F$ can be expressed at any order in $\hbar$ by $\delta M$, defined in 
Eq.\ \eqref{deltaM}, via the relation
\begin{equation}
\delta M_F = \frac{m}{p^2} \Tr(p\cdot\gamma\, \delta M)\; .
\end{equation}
Furthermore, from \eqs\eqref{A} and \eqref{SSS} we obtain
\begin{equation}
\left(p^2-m^2\right) \A^\mu=\hbar \, \delta M^\mu_A +\mathcal{O}(\hbar^2)
= \mathcal{O}(\hbar^2)\;, \label{Aonshell}
\end{equation}
with
\begin{equation}
\delta M^\mu_A = \Tr(\gamma^\mu\gamma^5 \delta M)\; ,
\end{equation}
which, as a quantity with axial-vector quantum numbers, 
is itself of order $\mathcal{O}(\hbar)$.
This shows that, under the assumptions adopted, the axial-vector component, 
unlike $\bar{\F}$, remains on the 
mass-shell at first order in $\hbar$ even in the presence of interactions.

The Boltzmann equations are derived from Eqs.\ \eqref{Vkin} and \eqref{Akin},
{}{using Eqs.\ (\ref{spinorthcoll}) and (\ref{newV}),} and read
{}{up to corrections of order $\mathcal{O}(\hbar^2)$}
\begin{subequations} \label{Fspintransport}
\begin{eqnarray}
 	p\cdot\partial\bar{\F}& =& m\, C_F \;,  \label{Ftransport} \\
  	p\cdot\partial\A^\mu&=& m\, C_A^\mu\; , \label{spintransport}
\end{eqnarray}
\end{subequations}
with $C_F=2C_\F$ and 
$C_A^\mu\equiv -\frac{1}{m}\epsilon^{\mu\nu\alpha\beta}p_\nu C_{\Sc\alpha\beta}$.  
From {}{Eqs.\ \eqref{wigboltz9}, \eqref{fbar}, 
and \eqref{Fspintransport} }
one finds 
\begin{subequations}\label{colcolcollocloc}
\begin{eqnarray}
C_F&=& \frac{1}{p^2}\Tr(p\cdot\gamma\, C)\; ,\\
C_A^\mu&=& \frac1m \Tr(\gamma^\mu\gamma^5 C)\; ,
\end{eqnarray}
\end{subequations}
which establishes the connection to $C$ given in Eq.\ \eqref{generalcollisionterm}.
Equations \eqref{colcolcollocloc} will be used to determine the right-hand sides of 
Eqs.~\eqref{Fspintransport}, which, together with 
{}{Eq.\ \eqref{spinorthcoll}}, form a closed system of equations for 
$\bar{\F}$ and $\A^\mu$, as will be explicitly shown in the following.

\section{Spin in phase space}
\label{secspph}

We now introduce spin as an additional variable in phase space
\cite{Zamanian:2010zz,Ekman:2017kxi,Ekman:2019vrv,Florkowski:2018fap,Bhadury:2020puc,Weickgenannt:2020aaf}. 
The advantage of this concept is that it immediately connects the first-principle 
quantum description to a ``classical'' description of spin, which can be used, e.g., for 
hydrodynamics \cite{Weickgenannt:2020aaf}. 
Furthermore, as we will see later, it combines the full dynamics of the Boltzmann-like 
equations \eqref{Fspintransport} into one scalar equation and provides a natural 
interpretation for the conservation laws 
and the collisional invariants \cite{Weickgenannt:2020aaf}. 

It is convenient to define the single-particle distribution function in the
phase space extended by the additional spin variable $\ms$ as
\beq
 \f(x,p,\ms)\equiv\frac12\left[\bar{\F}(x,p)-\ms\cdot\A(x,p)\right]\; . \label{spinproj}
\eeq
This definition holds at any order in $\hbar$.  We then introduce the covariant
integration measure
\begin{equation}
 \int dS(p)  \equiv \sqrt{\frac{p^2}{3 \pi^2}} \int d^4\ms\, \delta(\ms\cdot\ms+3)
 \delta(p\cdot \ms)\;,
\end{equation}
which has the properties
\begin{subequations}
\begin{eqnarray}
 \int dS(p) & = & 2\;, \\
 \int dS(p)\, \ms^\mu & = & 0 \;, \\
 \int dS(p)\, \ms^\mu \ms^\nu & = & - 2 \left( g^{\mu \nu} - \frac{p^\mu p^\nu}{p^2} \right)\;. 
 \label{dS_smu_snu}
\end{eqnarray}
\end{subequations}
Consequently,
\begin{subequations}
\begin{eqnarray}
 \bar{\F}(x,p)&=&\int dS(p)\, \f(x,p,\ms)\; , \\
 \A^\mu(x,p)&=& \int dS(p)\, \ms^\mu \f(x,p,\ms)\;.\label{A_int}
\end{eqnarray}
\end{subequations}
{Higher moments of $\f$ with respect to the variable $\ms$ can be also 
related to $\F$ and $\A^\mu$ and do not yield any further information.} 
From {}{Eqs.~\eqref{V}, \eqref{SSS}, and \eqref{newV} we obtain} 
relations for $ \V^\mu$ 
and $ \Sc^{\mu\nu}$, which
are valid {}{up to corrections of order $\mathcal{O}(\hbar^2)$}, 
\begin{align}
 \V^\mu(x,p)=& \int dS(p)\, \left(\frac1m p^\mu
 +\frac{\hbar}{2m} \partial_\nu  \Sigma_\ms^{\mu\nu}\right) \f(x,p,\ms)
 +\mathcal{O}(\hbar^2)\;,  \n \\
 \Sc^{\mu\nu}(x,p)=& \int dS(p)\, \left(\Sigma_\ms^{\mu\nu}
 +\frac{\hbar}{2m^2}\partial^{[\mu}p^{\nu]}\right)
 \f(x,p,\ms) +\mathcal{O}(\hbar^2)\; ,
 \label{V_int}
\end{align}
where we defined 
\begin{equation}
 \Sigma_{\ms}^{\mu\nu}\equiv -\frac1m \epsilon^{\mu\nu\alpha\beta}p_\alpha \ms_\beta\;.
\end{equation}
Similar as for the Wigner function, we can write a modified on-shell condition and the 
Boltzmann equation for the scalar distribution $\f$. Using 
{}{Eqs.\ \eqref{Fbaronshell} and \eqref{Aonshell}}, 
the on-shell condition is given by
\begin{equation}
\label{onshellfff}
 \left(p^2-m^2\right) \f(x,p,\ms) = \hbar \, \mathfrak{M}(x,p,\ms)
 + \mathcal{O}(\hbar^2)\; ,
\end{equation}
with 
\begin{equation}
 \mathfrak{M}(x,p,\ms) = \frac12 \left[\delta M_F(x,p) - \ms \cdot \delta M_A(x,p)\right]\;.
\end{equation}
In order to find a solution for Eq.\ \eqref{onshellfff} we employ the quasi-particle 
approximation, i.e.,
we assume that the distribution $\f$ is of the form
\begin{equation}
 \f(x,p,\ms)=m\delta(p^2-M^2) f(x,p,\ms)\;, \label{onshellsolution}
\end{equation}
where $f(x,p,\ms)$ is a function without singularity at 
$p^2=M^2 \equiv m^2+ \hbar\delta m^2$, with
$\delta m^2(x,p,\ms)$ being a correction to the
mass-shell condition for free particles arising from interactions. 
After Taylor-expanding the delta function to first order in $\hbar$ and assuming 
that $f(x,p,\ms)$ has no singularity at $p^2 = m^2$, i.e.,
$(p^2-m^2) \delta(p^2-m^2) f(x,p,\ms) =0$, we can relate 
$\delta m^2$ with $\mathfrak{M}$,
\begin{equation}
\hbar \,\mathfrak{M} (x,p,\ms)= \hbar \,\delta m^2 (x,p,\ms) \delta(p^2-m^2)mf(x,p,\ms)
+\mathcal{O}(\hbar^2)\;, 
\label{offshellfrominteractions}
\end{equation}
where we used $(p^2-m^2)\delta^\prime(p^2-m^2)=-\delta(p^2-m^2)$.
As a consequence of our assumption that spin degrees of freedom 
enter at first order, the $\ms$ dependence of $\hbar \delta m^2$ appears at least at 
$\mathcal{O}(\hbar^2)$. 

The Boltzmann equation for $\f$ is derived from 
{}{Eqs.\ \eqref{Fspintransport} and \eqref{spinproj}} and reads
\begin{equation}\label{Boltzmannnn}
 p\cdot\partial \, \f(x,p,\ms)=m\, \mathfrak{C} \;,
\end{equation}
where we introduced the collision kernel
\begin{equation}
\mathfrak{C}\equiv \frac{1}{2}(C_F-\ms \cdot C_A)\;. \label{spinphspcoll}
\end{equation}
As will be shown in the following, 
up to first order in $\hbar$ the collision term has the following structure,
\begin{equation} \label{fullcollterm}
	\mathfrak{C} =  \mathfrak{C}^{(0)}_{{l}}+\hbar \left\{ \mathfrak{C}^{(1)}_{{l}}
	+ \mathfrak{C}^{(1)}_{{nl}} \right\}
	 \equiv  \mathfrak{C}_{{l}} +\hbar \mathfrak{C}^{(1)}_{{nl}}\;.
\end{equation}
Here, local and nonlocal contributions are denoted by subscripts ${l}$ and ${nl}$,
respectively. As already mentioned, the zeroth-order contribution is 
purely local~\cite{Li:2019qkf}, while the first-order contribution
has both local and nonlocal parts. In the next sections, 
we will calculate the local and nonlocal collision terms explicitly.

\section{Local collisions}
\label{secloc}

In order to explicitly calculate the collision term, we follow Ref.~\cite{DeGroot:1980dk}. 
We first focus on the local part, i.e., the term $\sim \delta^{(4)}(q_i)$ in the second line 
of Eq.~\eqref{generalcollisionterm}. 
The matrix element of $\Phi$ appearing in this equation, with $\Phi$ given by 
Eq.~\eqref{phigen}, is calculated in App.~\ref{collkerapp}. 
The local contribution is thus obtained from Eq.~\eqref{offshellscatteringfinal} with 
$q_i=0$ \cite{DeGroot:1980dk}
\begin{align}
&(2\pi\hbar)^6\inlanglesm p_1,p_2;r_1,r_2|\Phi| p_1,p_2;s_1,s_2\inranglesm 
=\sum_{rs}u_r(p)\bar{u}_s(p)w_{r_1r_2s_1s_2}^{rs}(p_1,p_2,p) \; ,
\label{phizero}
\end{align}
with 
\begin{align}
w_{r_1r_2s_1s_2}^{rs}(p_1,p_2,p)& 
=2\delta(p^2-m^2)\Bigg\{ \sum_{r^\prime}\int dP^\prime 
\delta(p+p^\prime-p_1-p_2) \langle{p,p^\prime;r,r^\prime|t|p_1,p_2;s_1,s_2}\rangle
\langle{p_1,p_2;r_1,r_2|t^\dagger|p,p^\prime;s,r^\prime}\rangle \n\\
&+ \left[i\pi\hbar p^0\delta^{(3)}(\mathbf{p}-\mathbf{p}_1)
\left(\langle p,p_2;r,r_2|t|p,p_2;s_1,s_2\rangle 
\delta_{r_1s}-\langle p,p_2;r_1,r_2|t^\dagger|p,p_2;s,s_2\rangle \delta_{rs_1}\right)
+(1\leftrightarrow2) \right] 
\Bigg\}\;, \label{wwww}
\end{align}
where the symbol $(1\leftrightarrow2)$ denotes the exchange of the indices $1$ and $2$,
$dP\equiv d^4p\, \delta(p^2-m^2)$, and 
\begin{align}
&\langle{p,p^\prime;r,r^\prime|t|p_1,p_2;s_1,s_2}\rangle\equiv
-\sqrt{\frac{(2\pi\hbar)^{7}}{2}}\, \bar{u}_r(p) \outlanglesm p^\prime;r^\prime|:\rho(0):| 
p_1,p_2;s_1,s_2\inranglesm
 \label{turho}
\end{align}
is the conventional scattering amplitude due to the interaction $\rho$, which can be 
computed using standard 
techniques {}{from} 
quantum field theory \cite{DeGroot:1980dk,itzykson2012quantum}. 
We are now ready to calculate the local part of Eq.\ \eqref{fullcollterm}. To this end,
we insert Eq.~\eqref{phizero} into Eq.~\eqref{generalcollisionterm}, then 
Eq.~\eqref{generalcollisionterm} into  
Eqs.~\eqref{colcolcollocloc} and, finally, we plug Eqs.~\eqref{colcolcollocloc} into 
Eq.~\eqref{spinphspcoll}. 
In this way, the local part of the collision kernel is given by
\begin{equation}
\label{cloc1}
 \mathfrak{C}_l
 = \frac{1}{8m^4}\sum_{r_1,r_2,s_1,s_2}\int d^4 p_{1}d^4 p_2
 \sum_{r^\prime,s^\prime} h_{s^\prime r^\prime}(p,\ms)
 w_{r_1 r_2 s_1 s_2}^{r^\prime s^\prime}(p_1,p_2, p) 
 \prod_{j=1}^2 \bar{u}_{s_j}(p_j) W(x, p_j)u_{r_j}(p_j)\; ,
\end{equation}
where we have used
\begin{eqnarray}
p^\mu \delta_{sr} & \equiv &\frac{1}{2} \, \bar{u}_s(p)\gamma^\mu u_r(p)\; , \\
n^\mu_{sr}(p) &\equiv & \frac{1}{2m}\, \bar{u}_s(p)\gamma^5\gamma^\mu u_r(p)\; , 
\label{n_mu_rs}
\end{eqnarray}
and defined
\begin{equation}
h_{s r}(p,\ms) \equiv \delta_{s r}+\ms \cdot  n_{s r}(p)\; .
\end{equation} 
The factor $\delta(p^2-m^2)$ in \eq\eqref{wwww} shows that the local term is always 
on-shell.
This comes from the difference $G(p)-G^\star (p)=2\pi i \hbar^2 \delta (p^2 - m^2)$, with 
\begin{equation}
 G(p)=-\frac{\hbar^2}{p^2-m^2+i\epsilon p^0} \;, \label{gapp}
\end{equation}
which appears in  the first line of \eq\eqref{w1app} when we set $q_i=0$.
We now use the Clifford decomposition \eqref{dec} to write \eq\eqref{cloc1} as
\begin{align}  
 \mathfrak{C}_l ={}&\frac{1}{32 m^2}\sum_{r_1,r_2,s_1,s_2}\int d^4 p_{1}d^4p_2 
 \sum_{r^\prime,s^\prime}
 h_{s^\prime r^\prime}(p,\ms)w_{r_1 r_2 s_1 s_2}^{r^\prime s^\prime}(p_1,p_2,p) \n\\
  &\times\prod_{j=1}^2 \left[ \F(x,p_j)\delta_{s_jr_j}+\frac{1}{m} p \cdot \V(x,p_j)
  \delta_{s_jr_j}+n_{s_j r_j}(p_j)\cdot \A(x,p_j)
  +\frac12\Sigma^{\mu_j\nu_j}_{s_jr_j}(p_j) \Sc_{\mu_j\nu_j}(x,p_j)\right]\;,
  \label{Clocal_intermediate}
 \end{align} 
where we defined
\begin{equation}
\Sigma^{\mu\nu}_{rs} (p)\equiv \frac{1}{2m}\, \bar{u}_r(p)\sigma^{\mu\nu} u_s(p) 
= \frac{1}{m} \epsilon^{\mu \nu\alpha \beta} p_\alpha n_{rs \beta}(p)\;.
\label{Sigma_munnu_rs} 
\end{equation} 
Using Eqs.\ \eqref{A_int}, \eqref{V_int}, \eqref{onshellsolution}, the relations 
$p_\mu\Sigma^{\mu\nu}_\ms=p_\mu\Sigma_{sr}^{\mu\nu}=0$, and
\begin{equation}
 \Sigma_{sr}^{\mu\nu}\Sigma_{\ms\mu\nu}=2\frac{p^2}{m^2} \, \ms \cdot n_{sr} \;,
 \end{equation}
we can rewrite Eq.\ \eqref{Clocal_intermediate} in the form
\begin{align}  
 \mathfrak{C}_l ={}&\frac 18\sum_{r_1,r_2,s_1,s_2}\int d\Gamma_1\, d\Gamma_2\,
 \sum_{r^\prime,s^\prime} 
  h_{s^\prime r^\prime}(p,\ms) w_{r_1 r_2 s_1 s_2}^{r^\prime s^\prime}(p_1,p_2,p) 
  \prod_{j=1}^2 h_{s_j r_j}(p_j,\ms_j)  f(x,p_j,\ms_j) \; , \label{somanycollisionterms}
\end{align}
where 
\begin{equation}
\int d\Gamma \equiv \int d^4p\, \delta(p^2 - m^2) \int dS(p)\;.
\end{equation}
Plugging Eq.~\eqref{wwww} into \eqref{somanycollisionterms},  the collision term reduces to
\begin{equation}
\mC_l=\delta(p^2-m^2)\, \mC_{\osl}[f] \;,
\end{equation}
where
\begin{eqnarray}
\mathfrak{C}_{\osl}[f]&=& \frac14\sum_{r_1,r_2,s_1,s_2}\sum_{r,r^\prime,s}
\int d\Gamma_1\, d\Gamma_2\, dP^\prime\, h_{sr}(p,\ms)
\delta^{(4)}(p+p^\prime-p_1-p_2) 
 \n\\
  &&\times\langle{p,p^\prime;r,r^\prime|t|p_1,p_2;s_1,s_2}\rangle 
  \langle{p,p_2;r_1,r_2|t^\dagger|p,p^\prime;s,r^\prime}\rangle
  \prod_{j=1}^2  h_{s_jr_j}(p_j,\ms_j) \, f(x,p_j,\ms_j)\n\\
  & + & i\, \frac{\pi\hbar}{4}\sum_{r_2,s_1,s_2}\sum_{r,s}\int  d\Gamma_2\, dS_1(p)\, 
    h_{s_2r_2}(p_2,\ms_2)\, f(x,p,\ms_1)f(x,p_2,\ms_2) \n\\
  &&\times\left[ h_{sr}(p,\ms)  h_{s_1s}(p,\ms_1) \langle{p,p_2;r,r_2|t|p,p_2;s_1,s_2}\rangle 
   - h_{s_1s}(p,\ms)  h_{sr}(p,\ms_1) 
  \langle{p,p_2;r,r_2|t^\dagger|p,p_2;s_1,s_2}\rangle\right]
  \label{relabelledcollisions}
\end{eqnarray}
is the local collision term on the mass shell. Using the identity
\begin{equation}
\sum_{s^\prime} n^\mu_{rs^\prime}(p)n^\nu_{s^\prime s}(p)
=- \left(g^{\mu\nu}-\frac{p^\mu p^\nu}{m^2}\right)\delta_{rs}
+\frac im \epsilon^{\mu\nu\alpha\beta}p_\alpha n_{rs\beta}(p) \; , \label{spinmatrixproduct}
\end{equation}
we can simplify
\begin{eqnarray}
 \sum_s\left[ h_{sr}(p,\ms)h_{s_1s}(p,\ms_1)-h_{s_1s}(p,\ms)h_{sr}(p,\ms_1)
 \right]
& = &\ms_\mu\ms_{1\nu} \sum_s 
 \left[n^\mu_{sr}(p)n^\nu_{s_1s}(p)-n^\mu_{s_1s} (p)n^\nu_{sr}(p)\right]\n \\
& = & -i\,\frac 2m\ms_\mu\ms_{1\nu}\epsilon^{\mu\nu\alpha\beta}
p_\alpha n_{s_1r\beta}(p)\;,
\end{eqnarray}
to obtain
\begin{eqnarray}
 \mathfrak{C}_{\osl}[f]&=& \frac14\sum_{r_1,r_2,s_1,s_2}\sum_{r,r^\prime,s}
 \int d\Gamma_1\, d\Gamma_2\, 
 dP^\prime\,  h_{sr}(p,\ms)\delta^{(4)}(p+p^\prime-p_1-p_2) \n\\
  &&\times\langle{p,p^\prime;r,r^\prime|t|p_1,p_2;s_1,s_2}\rangle
  \langle{p,p_2;r_1,r_2|t^\dagger|p,p^\prime;s,r^\prime}\rangle
  \prod_{j=1}^2  h_{s_jr_j}(p_j,\ms_j) f(x,p_j,\ms_j)\n\\
  & + &i\, \frac{\pi\hbar}{8}\sum_{r_2,s_1,s_2}\sum_{r,s}
  \int  d\Gamma_2\, dS_1(p)\,    h_{s_2r_2}(p_2,\ms_2) \, f(x,p,\ms_1)f(x,p_2,\ms_2)\n\\
  &&\times\left[ h_{sr}(p,\ms)  h_{s_1s}(p,\ms_1) + h_{s_1s}(p,\ms)  h_{sr}(p,\ms_1) \right]
  \langle{p,p_2;r,r_2|t-t^\dagger|p,p_2;s_1,s_2}\rangle \n\\
  &+ &\, \frac{\pi\hbar}{4m}\sum_{r_2,s_1,s_2}\sum_{r}
  \int  d\Gamma_2\, dS_1(p)\,   h_{s_2r_2}(p_2,\ms_2) \, f(x,p,\ms_1)f(x,p_2,\ms_2)\n\\
  &&\times\,  \ms_\mu\ms_{1\nu}\epsilon^{\mu\nu\alpha\beta}p_\alpha n_{s_1r\beta}(p)
  \langle{p,p_2;r,r_2|t+t^\dagger|p,p_2;s_1,s_2}\rangle  \;.
\end{eqnarray}
The expression above can be further simplified by noting that the term involving 
the amplitude with the operator $t-t^\dagger$ is related to the first term through the 
optical theorem \cite{DeGroot:1980dk}
\begin{equation}
i\pi\hbar\langle p,p_1;r,r_1|t-t^\dagger|p,p_1;s,s_1\rangle 
= -\sum_{r^\prime,r_1^\prime} \int dP^\prime dP_1^\prime 
\langle p,p_1;r,r_1|t|p^\prime,p_1^\prime; r^\prime,r_1^\prime\rangle
\langle p^\prime,p_1^\prime; r^\prime, r_1^\prime|t^\dagger|p,p_1;s,s_1\rangle \; . 
\label{opttheo}
\end{equation}
Hence, the collision term is cast into the form
\begin{eqnarray}
 \mathfrak{C}_{\osl}[f] &=& \frac14\sum_{r_1,r_2,s_1,s_2}\sum_{r,r^\prime,s}
 \int d\Gamma_1\, d\Gamma_2\, dP^\prime\,  h_{sr}(p,\ms)
 \delta^{(4)}(p+p^\prime-p_1-p_2) \n\\
  & &\times\langle{p,p^\prime;r,r^\prime|t|p_1,p_2;s_1,s_2}\rangle
  \langle{p,p_2;r_1,r_2|t^\dagger|p,p^\prime;s,r^\prime}\rangle
  \prod_{j=1}^2  h_{s_jr_j}(p_j,\ms_j) f(x,p_j,\ms_j)\n\\
  & - &\frac18 \sum_{r_2,s_1,s_2}\sum_{r,s,r^\prime,r_1^\prime}
  \int  d\Gamma_2\, dP^\prime\, dP_1^\prime\, 
  dS_1(p)\,    h_{s_2r_2}(p_2,\ms_2) \, f(x,p,\ms_1)f(x,p_2,\ms_2)\n\\
 & &\times\left[ h_{sr}(p,\ms)  h_{s_1s}(p,\ms_1) 
 + h_{s_1s}(p,\ms)  h_{sr}(p,\ms_1) \right]
 \langle{p,p_2;r,r_2|t|p^\prime,p_1^\prime;r^\prime,r_1^\prime}\rangle 
  \langle{p^\prime,p_1^\prime;r^\prime,r_1^\prime|t^\dagger|p,p_2;s_1,s_2}\rangle \n\\
  &+& \frac{\pi \hbar}{4m}\sum_{r_2,s_1,s_2}\sum_{r}\int  d\Gamma_2\, dS_1(p)\,  
   h_{s_2r_2}(p_2,\ms_2) \, f(x,p,\ms_1)f(x,p_2,\ms_2) \n\\
&  &\times \,\ms_\mu\ms_{1\nu}\epsilon^{\mu\nu\alpha\beta}p_\alpha n_{s_1r\beta}(p)
  \langle{p,p_2;r,r_2|t+t^\dagger|p,p_2;s_1,s_2}\rangle \; .
\end{eqnarray}
In order to write the collision term in a compact form, we insert factors of one for the 
phase-space spin 
variable in the form $1=(1/2)\int dS(p)$ and obtain
\begin{equation} \label{localcoll_1}
\mathfrak{C}_{\osl}[f]\equiv  \mathfrak{C}_{\mathrm{p+s}}[f] 
+  \mathfrak{C}_{\mathrm{s}}[f]\;,
\end{equation}
with 
\begin{subequations}\label{looocalcoll}
\begin{eqnarray}
 \mathfrak{C}_{\mathrm{p+s}}[f]  & \equiv  &
\int d\Gamma_1\, d\Gamma_2\, d\Gamma^\prime \, dS_1^\prime(p)  \, \mathcal{W} 
 \big[  f(x,p_1,\ms_1)\big. f(x,p_2,\ms_2)\big.
 - f(x,p,\ms_1^\prime)f(x,p^\prime,\ms^\prime)\big]\;,\label{Cp+s} \\
\mathfrak{C}_{\mathrm{s}}[f] &\equiv & \int  d \Gamma_2\, dS_1(p)\,
\mathfrak{W}\, f(x,p,\ms_1)f(x,p_2,\ms_2)\;, \label{Cs}
\end{eqnarray}
\end{subequations}
where
\begin{eqnarray}
  \mathcal{W} &\equiv &\frac{1}{32} \sum_{s,r,s_1^{\prime}} \left[
  h_{s s_1^\prime} (p,\ms_1^\prime) h_{s_1^\prime r}(p,\ms) +
  h_{s s_1^\prime}(p,\ms) h_{s_1^\prime r} (p,\ms_1^\prime)\right] 
  \sum_{s',r',s_1,s_2,r_1,r_2} h_{s^\prime r^\prime}(p^\prime, \ms^\prime) \,
  h_{s_1 r_1}(p_1, \ms_1)  \, h_{s_2 r_2}(p_2, \ms_2) \n \\
    &&\times \langle{p,p^\prime;r,r^\prime|t|p_1,p_2;s_1,s_2}\rangle
  \langle{p_1,p_2;r_1,r_2|t^\dagger|p,p^\prime;s,s^\prime}\rangle
   \delta^{(4)}(p+p^\prime-p_1-p_2)\; \label{local_col_GLW}
\end{eqnarray}
and
\begin{eqnarray}
 \mathfrak{W}&\equiv& \frac{\pi\hbar}{4m} \sum_{s_1,s_2,r,r_2}
 \epsilon_{\mu\nu\alpha\beta} \ms^\mu \ms_1^{\nu} p^\alpha
 n_{s_1r}^{\beta}(p)\, h_{s_2r_2} (p_2, \ms_2)
 \langle{p,p_2;r,r_2|t+t^\dagger|p,p_2;s_1,s_2}\rangle \;.\label{mathw}
\end{eqnarray}
The term $\mathfrak{C}_{\mathrm{p+s}}[f]$ in Eq.\ \eqref{localcoll_1}
describes momentum- and spin-exchange interactions,
while the term $\mathfrak{C}_{\mathrm{s}}[f]$ {}{corresponds to} 
spin exchange without momentum exchange.
If the distribution functions do not depend on the spin variables, i.e., $f(x,p,\ms)
\equiv f(x,p)$, we recover the collision term familiar from the Boltzmann equation, where 
averaging and summation over spins is done directly in the cross section. 
We do not obtain Pauli--blocking factors because of the low-density 
approximation~\cite{DeGroot:1980dk}.
Moreover, this implies that, in equilibrium, only Boltzmann instead of 
Fermi--Dirac distributions will appear. 
If the distribution functions depend on spin, the two terms on the
right-hand side of Eq.~\eqref{localcoll_1} require further discussion.
Considering $\mathfrak{C}_{\mathrm{p+s}}[f]$, the term 
$\sim f(x,p_1,\ms_1)f(x,p_2,\ms_2)$
has the form of a gain term for particles with momentum $p$ and spin $\ms$, while
$\sim f(x,p,\ms_1^\prime)f(x,p^\prime,\ms^\prime)$ does not have an obvious 
interpretation as a loss
term, because the spin variable is $\ms_1^\prime$, not $\ms$.

However, it is possible to define a new distribution function and a new collision term
such that we recover the standard interpretation of gain and loss terms in the latter 
without changing the physics.
The underlying idea is that, since the phase-space spin variable is not observable, 
physical quantities are only obtained after integrating
over the former, see, e.g., Eq.~\eqref{A_int}. Therefore, we seek to replace
the distribution function $f (x,p,\ms)$ by another distribution function 
$\tilde{f}(x,p,\ms)$ and likewise the collision term
$\mC[f]$ by $\tilde{\mC}[\tilde{f}]$, such that 
\begin{subequations} \label{wep}
\begin{eqnarray}
 p \cdot \partial \, \tilde{f}(x,p,\ms)&=&\tilde{\mC}[\tilde{f}]\;, \label{prop1}\\
 \int dS(p)\, b\, \tilde{Q}(x,p,\ms)&=& \int dS(p)\, b\, Q(x,p,\ms) \label{prop2}
\end{eqnarray}
\end{subequations}
is fulfilled for $Q\in \{f,\mC[f]\}$, $\tilde{Q}\in \{\tilde{f},\tilde{\mC}[\tilde{f}]\}$,
$b\in \{1,\ms^\mu\}$. In other words, after integration over the spin variable,
the new distribution function $\tilde{f}(x,p,\ms)$ and the collision term 
$\tilde{\mC}[\tilde{f}]$ are
equivalent to the old ones. Moreover, they fulfill the same Boltzmann equation.
Consequently, the new quantities 
$\tilde{f}(x,p,\ms)$ and $\tilde{\mC}[\tilde{f}]$ describe the same physics as the old ones.
Equations~\eqref{wep} constitute a ``weak equivalence
principle'', stating that $f$ and $\tilde{f}$ formally obey the same equation of motion and
give identical results when integrating over the spin variable.

We now want to derive the collision term $\tilde{\mC}_{\textrm{p+s}}[\tilde{f}]$,
which satisfies the weak equivalence principle.
The ultimate goal is to modify Eq.\ \eqref{Cp+s} and the first line of 
\eq\eqref{local_col_GLW} in such a way that $\ms_1$ is replaced by $\ms$ and 
the integration over $dS_1^\prime(p)$ disappears, so that one obtains a collision term 
which has a standard gain and loss term. According to Eq.~\eqref{spinphspcoll}, we will 
show this separately for the (p+s) parts of $C_F$ and $C_A^\mu$, respectively. 
The calculation for $C_F$ is actually straightforward, so we only present
the somewhat more complicated case of $C_A^\mu$.
Considering $\int dS(p) \, \ms^\alpha\, \mC_{\textrm{p+s}}[f]$, one encounters 
a term of the form
\begin{align}
\frac12\sum_{s_1^\prime}\int dS(p) dS_1^\prime(p)\, \ms^\alpha \ms_\mu 
\ms_1^{\prime\beta} \ms_{1\nu}^\prime  [n^\mu_{s_1^\prime r} (p) n_{ss_1^\prime}^\nu(p)
+n^\mu_{ss_1^\prime} (p) n_{s_1^\prime r}^\nu(p) ]
={}&- \int dS(p) dS_1^\prime(p) \ms^\alpha \ms_\mu
\ms_1^{\prime\beta}\ms_{1\nu}^{\prime}
\left(g^{\mu\nu}-\frac{p^\mu p^\nu}{p^2}\right)\delta_{rs}\n\\
 ={}& 2 \int dS(p)\, \ms^\alpha \ms^\beta\delta_{rs}
 = - 4\left(g^{\alpha\beta}-\frac{p^\alpha p^\beta}{p^2}\right)\delta_{rs}\;,
 \label{caweak}
\end{align}
where we used Eq.~\eqref{spinmatrixproduct} in the first and Eq.~\eqref{dS_smu_snu} 
in the last step.
With this relation, and similar ones for the other terms, we see that, after integration
over $dS(p)$, the replacement
\begin{align}
\sum_{s_1^{\prime}} \int dS_1^\prime(p) \left[ h_{s s_1^\prime} (p,\ms_1^\prime) 
h_{s_1^\prime r}(p,\ms) 
+ h_{s s_1^\prime}(p,\ms) h_{s_1^\prime r} (p,\ms_1^\prime)\right] 
&\longrightarrow  4 h_{sr}(p,\ms) \label{specialweak}
\end{align}
fulfills the weak equivalence principle.

Furthermore, by definition, Eq.~\eqref{spinproj}, $f$ is linear in $\ms$ and, 
since $\A^{(0)\mu} =0$, 
the $\ms$ dependence enters only at first order in $\hbar$, i.e., $f=f(\hbar\ms)$. 
We assume that both $f$ and $\tilde{f}$ can be Taylor-expanded in terms of $\hbar \ms$. 
Inserting this Taylor expansion on the left- and right-hand sides of Eq.\ (\ref{prop2}),
we conclude that, to order $\mathcal{O}(\hbar)$, the only choice for
$\tilde{f}$ is $\tilde{f}\equiv f$, with deviations entering at order $\mathcal{O}(\hbar^2)$. 
With Eq.\ (\ref{specialweak}) we then find that, with
the relations $\int dS(p)\, \ms^\mu=0$ and $\int dS(p)\, \ms^\mu \ms^\nu \ms^\lambda=0$, 
the choice
\begin{eqnarray}
\tilde{\mC}_{\mathrm{p+s}}[f] &\equiv  &
 \int d\Gamma_1\, d\Gamma_2\, d\Gamma^\prime\,
 \widetilde{\mathcal{W}} \big[ f(x,p_1,\ms_1)f(x,p_2,\ms_2) 
 -f(x,p,\ms)f(x,p^\prime,\ms^\prime)\big]\;,
 \label{localcollafter}
\end{eqnarray}
with
\begin{eqnarray}
\widetilde{\mathcal{W}}&\equiv& \delta^{(4)}(p+p^\prime-p_1-p_2)\,
 \frac{1}{8} \sum_{s,r}   h_{s r} (p,\ms)  
 \sum_{s',r',s_1,s_2,r_1,r_2} h_{s^\prime r^\prime}(p^\prime, \ms^\prime) \,  
 h_{s_1 r_1}(p_1, \ms_1)
  \, h_{s_2 r_2}(p_2, \ms_2) \n \\
    &&\times \langle{p,p^\prime;r,r^\prime|t|p_1,p_2;s_1,s_2}\rangle
  \langle{p_1,p_2;r_1,r_2|t^\dagger|p,p^\prime;s,s^\prime}\rangle \;,
  \label{local_col_GLW_after}
\end{eqnarray}
satisfies the weak equivalence principle \eqref{wep} up to $\mathcal{O}(\hbar)$.

Let us now focus on $\mC_{\mathrm{s}}[f]$. We will argue that this term has already 
the expected structure with gain and loss terms. In order to see this, let us first note 
that $\mC_{\mathrm{s}}[f]$ corresponds to collisions where the momentum of each particle
is conserved, but the spin can change: 
$(p,\ms_1),(p_2,\ms_2)\rightarrow (p,\ms),(p_2,\ms^\prime)$
\cite{DeGroot:1980dk}.
Here, the distribution functions $f(x,p,\cdot)$ and $f(x,p^\prime,\cdot)$ describe 
the particles before \textit{and} after the collision, which means that they contribute to 
\textit{both} the gain and the
loss term. We see from Eq.~\eqref{mathw} that the interchange of $\ms^\mu$ and 
$\ms^\nu_1$ flips the sign of $\mathfrak{W}$. This means that a net gain of particles with
($p,\ms$) corresponds to a net loss of particles with $(p,\ms_1)$. Thus, 
$\mC_{\mathrm{s}}[f]$ contains both gain and loss terms.

\section{Nonlocal collisions}
\label{secnon}

In order to calculate the nonlocal collision term, we focus on the second term in 
the second line of Eq.~\eqref{generalcollisionterm}. Note that the 
$\partial^\mu_{q_j} \delta^{(4)}(q_j)$ term implies that the 
momentum $p^\mu$ of the Wigner function is not on-shell anymore, which is in contrast 
to the local term. 
In fact, when we integrate by parts, the nonlocal kernel in 
\eq\eqref{generalcollisionterm} can be divided into two terms
\begin{equation}
\label{cnl11}
\mC_{nl}^{(1)}=\mC_{nl,1}^{(1)}+\mC_{nl,2}^{(1)} \;.
\end{equation}
In the first term the $q_j$-derivative acts on the spinors, i.e.,
\begin{align}
 \mC_{nl,1}^{(1)} = {}&\frac{ i}{8 m^4}\sum_{r_1,r_2,s_1,s_2}
\int d^4 p_{1}d^4 p_{2}d^4 q_{1}d^4 q_{2} \, \delta^{(4)}(q_1)\delta^{(4)}(q_2)
 \n\\
&\times \Tr\left[\left(\frac{1}{p^2}\, p\cdot\gamma-\frac1m\ms \cdot \gamma \, 
\gamma^5\right) (2\pi\hbar)^6\inlangle{p_1-\frac{q_1}{2},p_2-\frac{q_2}{2};r_1,r_2\Big|
\Phi(p)\Big|p_1+\frac{q_1}{2}, p_2+\frac{q_2}{2};s_1,s_2}\inrangle\right]\n\\
  &\times\left\{ \bar{u}_{s_1}\left(p_1+\frac{q_1}{2}\right) 
  W(x, p_1)u_{r_1}\left(p_1-\frac{q_1}{2}\right)\partial_{q_2}^\mu 
  \left[\bar{u}_{s_2}\left(p_2+\frac{q_2}{2}\right) 
  \partial_{\mu}W(x,p_2) u_{r_2}\left(p_2-\frac{q_2}{2}\right)\right] \right.\n\\
&+ \left. \partial_{q_1}^\mu \left[ \bar{u}_{s_1}\left(p_1+\frac{q_1}{2}\right) 
\partial_{\mu}W(x, p_1) 
u_{r_1}\left(p_1-\frac{q_1}{2}\right) \right] \bar{u}_{s_2}\left(p_2+\frac{q_2}{2}\right) 
W(x,p_2)u_{r_2}\left(p_2-\frac{q_2}{2}\right) \right\} \;, \label{cnlfirst}
\end{align}
which is on-shell because of $\delta^{(4)}(q_1)\delta^{(4)}(q_2)$,  i.e.,
\begin{equation}
\mC_{nl,1}^{(1)} =\delta(p^2-m^2)\,\mC_{\os,nl,1}^{(1)}\;.
\end{equation}
In the second term in Eq.\ (\ref{cnl11})  
the $q_j$-derivative acts on the matrix element of $\Phi$, i.e.,
\begin{align}
 \mC_{nl,2}^{(1)} = {}& \frac{i}{8 m^4}\sum_{r_1,r_2,s_1,s_2}
\int d^4 p_{1}d^4 p_{2}d^4 q_{1}d^4 q_{2} \, \delta^{(4)}(q_1)\delta^{(4)}(q_2)\n\\
  &\times\left\{ \bar{u}_{s_1}\left(p_1+\frac{q_1}{2}\right) W(x, p_1)u_{r_1}
  \left(p_1-\frac{q_1}{2}\right) 
  \bar{u}_{s_2}\left(p_2+\frac{q_2}{2}\right) \partial_{\mu}W(x,p_2)u_{r_2}
  \left(p_2-\frac{q_2}{2}\right) 
  \partial_{q_2}^\mu\right.\n\\
&+ \left.  \bar{u}_{s_1}\left(p_1+\frac{q_1}{2}\right) \left[ \partial_{\mu}W(x, p_1) \right]
u_{r_1}\left(p_1-\frac{q_1}{2}\right) 
\bar{u}_{s_2}\left(p_2+\frac{q_2}{2}\right) W(x,p_2)u_{r_2}\left(p_2-\frac{q_2}{2}\right) 
\partial_{q_1}^\mu\right\}\n\\
&\times\,  \Tr\left[\left(\frac{1}{p^2}\, p\cdot\gamma-\frac1m \ms \cdot \gamma \, 
\gamma^5\right) (2\pi\hbar)^6   
\inlangle{p_1-\frac{q_1}{2},p_2-\frac{q_2}{2};r_1,r_2\Big|\Phi(p)\Big|
p_1+\frac{q_1}{2}, p_2+\frac{q_2}{2};s_1,s_2}\inrangle\right] \;,
\label{cnlsecond}
\end{align}
which in general contains off-shell parts, cf.\ App.~\ref{non_loc_coll_calc}. 
Note that the factor $(2\pi\hbar)^6$ in front of the matrix element in 
Eqs.\ (\ref{cnlfirst}), (\ref{cnlsecond}) is
part of the normalization of the latter, cf.\ Eq.\ (\ref{phizero}), and does not participate 
in the $\hbar$-counting,
which is obvious since it is not accompanied by any gradient $\partial_\mu$.
As discussed in relation to \eq\eqref{approx1}, in our expansion scheme the zeroth-order 
distribution function is identified with the equilibrium 
distribution. Off-equilibrium contributions are at least of first order in gradients 
and will thus enter the nonlocal 
collision term only at higher order. 
Thus, in the following discussion we will always assume that
the distribution functions in the nonlocal collision term are identical to the 
zeroth-order equilibrium distribution.

In App.~\ref{non_loc_coll_calc}, it is shown that $\mC_{nl,2}^{(1)}$ can 
be divided into an on-shell and an off-shell part,
\begin{align}
\mC_{nl,2}^{(1)}={}&\mC_{\text{off-shell}}^{(1)} +\delta(p^2-m^2) \,
\mC_{\text{on-shell},2}^{(1)} \n\\
={}&\mC_{\text{off-shell}}^{(1)}+ \delta(p^2-m^2)
\left( \mC_{\os,2,1}^{(1)}+ \mC_{\os,2,2}^{(1)}\right) \;,
\end{align}
and it is proved that the off-shell contribution $\mC_{\text{off-shell}}^{(1)}$ cancels the 
off-shell part of the left-hand side of the Boltzmann equation \eqref{Boltzmannnn} 
when substituting \eq\eqref{onshellsolution}. 
Furthermore, in App.~\ref{non_loc_coll_calc}, we show that $\mC_{\os,2,1}^{(1)}=0$, 
if one inserts the 
zeroth-order equilibrium distribution [see \eq\eqref{ceq0}].
The explicit expression for $\mC_{\os,2,2}^{(1)}$ contains momentum derivatives of 
matrix elements and is 
computed in App.~\ref{non_loc_coll_calc}, see \eq\eqref{cmomder}. 
This term is neglected as we assume the 
scattering amplitude to be constant over scales of order of the interaction range 
defining the scattering 
nonlocality. This is consistent with the low-density approximation, 
see e.g.\ Ref.\ \cite{Abrikosov1975}.
Therefore the Boltzmann equation contains only on-shell contributions and 
can be written in terms of the distribution function $f(x,p,\ms)$ as
\begin{equation}
 \delta(p^2-m^2)\, p \cdot \partial f(x,p,\ms)=\delta(p^2-m^2)\, 
 \mathfrak{C}_{\text{on-shell}}[f]\;, \label{onshellcolll}
\end{equation}
with
\begin{equation}
\label{coll123}
 \mC_{\text{on-shell}}[f]\equiv  \mC_{\osl}[f]+\hbar\, \mC_{\os,nl,1}^{(1)}[f] \;,
\end{equation}
where $ \mC_{\osl}$ is the local term calculated in the previous section.

We are now ready to calculate $\mC_{nl,1}^{(1)}$. We note that, since in our scheme 
the lowest-order contribution to $\A^\mu$ and 
$\Sc^{\mu\nu}$ is of first order in gradients, these terms can be neglected in the 
nonlocal collision term, 
as the total contribution would be of second order. Using the spinor identities 
Eqs.\ \eqref{uu}, the relevant 
terms to compute in \eq\eqref{cnlfirst}  are of the form
\begin{eqnarray} \label{zzzzzzzzzzz}
&&i \partial_{q_j}^\mu\left[ \bar{u}_{s_j}\left(p_j+\frac{q_j}{2}\right)
\partial_{\mu}W(x,p_j)u_{r_j}\left(p_j-\frac{q_j}{2}\right)\right]_{q_j=0} \n\\
&=& i\left[\partial_{q_j}^\mu\bar{u}_{s_j}\left(p_j+\frac{q_j}{2}\right)
u_{r_j}\left(p_j-\frac{q_j}{2}\right)
\partial_{\mu} \F^{(0)}(x,p_j)+\partial_{q_j}^\mu\bar{u}_{s_j}\left(p_j+\frac{q_j}{2}\right)
\gamma^\alpha 
u_{r_j}\left(p_j-\frac{q_j}{2}\right)\partial_{\mu}\V^{(0)}_\alpha(x,p_j) 
\right]_{{}{q_j=0}}\n\\
  &=&\frac{1}{p_j^0+m} \, p_{j\nu}\Sigma^{\mu\nu}_{s_jr_j}(p_\star)\partial_{\mu}
  f^{(0)}(x,p_j)\n\\
 &=&  \frac{1}{p_j^0+m}\, \left[ \mathbf{p}_{j}\times\mathbf{n}_{s_jr_j}(p_j)\right]
 \cdot\boldsymbol{\nabla}f^{(0)}(x,p_j)\n\\
 &=& - \frac{1}{2 (p_j^0+m)} \int dS_j(p_j)\, 
 h_{s_jr_j}(p_j,\ms_j)(\mathbf{p}_{j}\times\boldsymbol{\ms}_j)
 \cdot\boldsymbol{\nabla}f^{(0)}(x,p_j)\;,
\end{eqnarray}
where $p^\mu_\star\equiv (m,\mathbf{0})$ is the four-momentum in the 
rest frame of the particle.
Using the result \eqref{zzzzzzzzzzz}, defining the space-time shift
\begin{equation}
\label{deltanon}
\Delta^\mu\equiv -\frac{\hbar}{2m(p\cdot\hat{t}+m)}\, 
\epsilon^{\mu\nu\alpha\beta}p_\nu \hat{t}_\alpha \ms_{\beta}\;,
\end{equation}
where $\hat{t}^\mu$ is the time-like unit vector which is $(1,\boldsymbol{0})$ 
in the frame where $p^\mu$ is measured, and applying similar steps as in the derivation of
Eq.~\eqref{looocalcoll}, we find that \eq\eqref{cnlfirst} becomes
\begin{eqnarray}
\hbar\, \mathfrak{C}^{(1)}_{\os,nl,1}[f] &=& \int d\Gamma_1\, d\Gamma_2\, 
d\Gamma^\prime\, dS_1^\prime(p)\, 
\mathcal{W} \n \\
& & \times \bigg[f(x,p_2,\ms_2)\Delta_1\cdot\partial f(x,p_1,\ms_1)
+f(x,p_1,\ms_1)\Delta_2\cdot\partial f(x,p_2,\ms_2) \bigg. \n \\
&& \hspace*{0.2cm} -f(x,p^\prime,\ms^\prime)\Delta_1^\prime\cdot\partial 
f(x,p,\ms_1^\prime)
-f(x,p,\ms_1^\prime)\Delta^\prime\cdot \partial f(x,p^\prime,\ms^\prime)\bigg]\n\\
 &+ & \frac{\pi \hbar}{4m}\sum_{r,r_2,s_1,s_2}\int  d\Gamma_2\, dS_1(p)\,   
 h_{s_2r_2}(p_2,\ms_2) \,\ms_\mu\ms_{1\nu}\epsilon^{\mu\nu\alpha\beta}
 p_\alpha n_{s_1r\beta}
  \langle{p,p_2;r,r_2|t+t^\dagger|p,p_2;s_1,s_2}\rangle \n\\
  && \times
  [f(x,p,\ms_1)\Delta_2\cdot\partial f(x,p_2,\ms_2)
  +f(x,p_2,\ms_2)\Delta_1\cdot\partial f(x,p,\ms_1)]\;.
\label{firstcontributiontofirstordercollision}
\end{eqnarray}
We now observe that  $\Delta\cdot\partial f(x,p,\ms)$ is the first-order contribution of the 
Taylor expansion of $f(x+\Delta,p,\ms)$. Hence, after applying the weak equivalence 
principle \eqref{wep}  to Eq.~\eqref{firstcontributiontofirstordercollision}, 
we can summarize the total collision term up to first order as
\begin{eqnarray}
\tilde{\mC}_{\os}[f] &  =& \int d\Gamma_1 d\Gamma_2 d\Gamma^\prime\,    
\widetilde{\mathcal{W}}\,  
[f(x+\Delta_1,p_1,\ms_1)f(x+\Delta_2,p_2,\ms_2)
-f(x+\Delta,p,\ms)f(x+\Delta^\prime,p^\prime,\ms^\prime)]\n\\
 &+& \int  d\Gamma_2 \, dS_1(p)\,\mathfrak{W}f(x+\Delta_1,p,\ms_1)
 f(x+\Delta_2,p_2,\ms_2)\;. \label{finalcollisionterm} 
\end{eqnarray}
The interpretation of Eq.~\eqref{finalcollisionterm} is the following: 
Incoming and outgoing particles {}{are dislocated from
the geometric center $x$ of the collision} by a space-like distance 
$\Delta^\mu$. This leads to a 
finite difference between the incoming and outgoing orbital angular momentum. 
This angular momentum is 
converted into spin polarization through a collision, which leads to the alignment of 
spin with the direction of 
vorticity discussed in Ref.~\cite{Weickgenannt:2020aaf}. 

We remark that the nonlocality in Eq.~\eqref{finalcollisionterm} should be distinguished 
from the nonlocality of a 
collision due to the side-jump effect in the massless case without interactions as it 
was discussed in Refs.~\cite{Chen:2015gta,Stone:2014fja}. 
The latter arises due to the anomalous Lorentz transformation of the center of inertia 
of massless particles 
\cite{Pryce:1948pf,Lorce:2018zpf,Speranza:2020ilk}. If a collision of massless particles 
is local in one 
reference frame, it will in general be nonlocal in a boosted reference frame, as the 
transformation behavior of the center of 
inertia leads to a position shift after the collision. On the other hand, for massive particles, 
it is always possible to define a space-time 4-vector associated with the center of mass, 
which properly transforms as a Lorentz vector as long as only local collisions are 
considered \cite{Weickgenannt:2020aaf,Speranza:2020ilk}. In other words, if the collision 
is local in one reference frame, it will stay local in all 
other reference frames. Hence, massive particles with local collisions will not experience 
any side-jump effect. 
The nonlocality in Eq.~\eqref{finalcollisionterm} is thus a nonlocality (in the sense 
of a finite impact parameter) in all reference frames, and, therefore, there is 
no ``no-jump frame''. In the massless case, this kind of nonlocality could be considered 
on top of the side-jump effect by introducing a collision 
which is not local even in the center-of-momentum frame. For a recent review about the 
difference between the centers of inertia and of mass and their connection to field theory 
see Ref.\ \cite{Speranza:2020ilk}.

\section{Equilibrium}

In order to find the conditions necessary to reach equilibrium,
we consider the standard form of the local equilibrium distribution 
function~\cite{Becattini:2013fla,Florkowski:2017ruc,Florkowski:2018fap}
\begin{equation}\label{f_eq}
 f_{\text{eq}}(x,p,\ms)=\frac{1}{(2\pi\hbar)^3}\exp\left[-\beta (x)\cdot p
 +\frac\hbar4 \Omega_{\mu\nu}(x)\Sigma_\ms^{\mu\nu}\right]\;.
\end{equation}
The exponent in \eq\eqref{f_eq} is a linear combination of the conserved quantities, 
which are momentum and total angular
momentum, where the Lagrange multipliers $\beta^\mu(x)=u^\mu(x)/T(x)$ and
$\Omega^{\mu\nu}(x)$ have the interpretation of fluid velocity over temperature and 
spin potential, respectively~\cite{Florkowski:2017ruc,Becattini:2018duy}. 
Here, we absorbed the orbital part of the angular momentum into the definition of
$\beta^\mu(x)$~\cite{Becattini:2013fla} and for the sake of simplicity considered the 
case of zero chemical potential, which corresponds to uncharged particles.
We now insert Eq.~\eqref{f_eq} into Eq.\eqref{finalcollisionterm} and obtain 
after expanding up to first order in $\hbar$
\begin{eqnarray}
 \tilde{\mathfrak{C}}_{\os}[f_{\text{eq}}] 
 &= & - \int d\Gamma^\prime d\Gamma_1 d\Gamma_2  \,
 \widetilde{\mathcal{W}} \,  e^{-\beta\cdot(p_1+p_2)}\n\\
 & &\times  \left[\partial_\mu\beta_\nu \frac{}{}
 \left(\Delta_1^\mu p_1^\nu+\Delta_2^\mu p_2^\nu-\Delta^\mu p^\nu-\Delta^{\prime\mu}
 p^{\prime\nu} \right)
  - \frac\hbar4\Omega_{\mu\nu}\left(\Sigma_{\ms_1}^{\mu\nu}
 +\Sigma_{\ms_2}^{\mu\nu}
 -\Sigma_{\ms}^{\mu\nu}-\Sigma_{\ms^\prime}^{\mu\nu}\right) \right]\n\\
 && - \int  d\Gamma_2\,  dS_1(p) dS^\prime(p_2)\, \mathfrak{W} \, 
 e^{-\beta\cdot(p+p_2)}\n \\
 & & \times \left\{\partial_\mu \beta_\nu \left[(\Delta_1^\mu - \Delta^\mu) p^\nu 
 + (\Delta_2^{\mu} - \Delta^{\prime \mu}) p_2^{\nu} \right] 
 -\frac\hbar4 \Omega_{\mu\nu}(\Sigma_{\ms_1}^{\mu\nu}
   +\Sigma_{\ms_2}^{\mu\nu}
   -  \Sigma_{\ms}^{\mu\nu}-\Sigma_{\ms^\prime}^{\mu\nu})\right\}\;.
    \label{colleqq}
\end{eqnarray}
Here, we used that the zeroth-order contribution to the collision term 
vanishes for the distribution function \eqref{f_eq}.
As the orbital angular momentum tensor of the
particle with $(p,\ms)$ is given by $L^{\mu\nu}=\Delta^{[\mu}p^{\nu]}$, the parentheses
in the second and {}{fourth} line can be interpreted as the balance of 
orbital angular momentum in the respective collision.

Defining the total angular momentum
$J^{\mu\nu}=L^{\mu\nu}+\frac\hbar2\Sigma_\ms^{\mu\nu}$ of the particle, 
which is assumed to be conserved in a collision,
$J^{\mu\nu}+J^{\prime\mu\nu}=J_1^{\mu\nu}+J_2^{\mu\nu}$,
the conditions for the vanishing of the collision term are for any 
$\widetilde{\mathcal{W}}$, $\mathfrak{W}$ given by
\begin{align}
\partial_{\mu}\beta_{\nu}+\partial_{\nu}\beta_{\mu}&=0\;, \label{cond1} \\
\Omega_{\mu\nu}=\varpi_{\mu\nu}&\equiv-\frac12 \partial_{[\mu}\beta_{\nu]} 
= \mathrm{const.}\;.
\label{cond2}
\end{align}
This corresponds to global (and not just local) equilibrium.

We remark that the conditions for global equilibrium given in \eqs\eqref{cond1} and 
\eqref{cond2} are necessary for the collision term to vanish if the standard ansatz for the 
equilibrium distribution function in \eq\eqref{f_eq} is used. In other words the standard 
concept of local equilibrium cannot be adopted when {}{both} 
spin effects and nonlocal collisions are considered. This result can be understood by 
looking at the ordering of scales. The usual way to derive hydrodynamics 
{}{from} kinetic theory is by assuming that there is a clear separation 
between some microscopic scale associated to the mean free path 
{}{$\ell_\text{mfp}$}, and a 
macroscopic scale associated to the hydrodynamic gradients 
{}{$L_\text{hydro}$}. One can 
then define the so-called Knudsen number {}{Kn
 \begin{equation} \label{Knudsen}
\text{Kn}\equiv \frac{\ell_\text{mfp}}{L_\text{hydro}}\;,
 \end{equation}
and require that Kn $\ll 1$.}
The expansion near equilibrium as defined in \eq\eqref{approx1} is a measure of 
how important {}{dissipative} effects are and can be associated to an 
expansion in the Knudsen number. Qualitatively, we can write 
\begin{equation}
\label{expeq}
\frac{\delta f}{f_{\text{eq}}} \sim \mathcal{O}(\text{Kn}) \,,
\end{equation}
where $\delta f$ is a function which describes deviations {of both the scalar and axial-vector part of the Wigner function} from equilibrium. For systems 
made of particles with spin, in addition to the expansion in \eq\eqref{expeq}, we introduce 
an expansion in powers of $\hbar$. In our framework, spin effects and scattering 
nonlocality are treated {}{as being of} 
the same order and it is natural to consider $\Delta$ as a new 
scale of the system. We can now relate {}{a new parameter $\kappa$ to 
the $\hbar$-expansion} {of the Wigner function}, defined as
\begin{equation}
\kappa \equiv \frac{\Delta}{L_\text{hydro}} \sim \frac{\hbar f^{(1)}}{f^{(0)}}\; .
\end{equation} 
The relation between $\kappa$ and the Knudsen number is given by
\begin{equation}
\kappa = \frac{\Delta}{\ell_\text{mfp}} \text{Kn}\,.
\end{equation}
In order for {}{the assumption of} molecular chaos to hold, and hence for 
particles to be considered as free between scatterings, we require that 
$\Delta \lesssim \ell_\text{mfp}$, implying 
\begin{equation}
\label{condkappa}
\kappa \lesssim \text{Kn} \;.
\end{equation}
The physical implication of this condition is that  {a local-equilibrium description of a fluid with spin and nonlocal collisions in kinetic theory would be inconsistent with the power counting. } In fact, if we consider spin as 
{}{of}
first order in $\kappa$, we cannot neglect dissipative effects at first order in Kn. 
For related discussions see also 
{}{Refs.}~\cite{Montenegro:2017rbu,Montenegro:2020paq}.

We can also express the condition in \eq\eqref{condkappa} in terms of the properties 
of the system. {}{To this end},
consider $\sigma$ to be a cross section and {}{$n$ 
the particle density. The geometric area given by the cross section, $\sigma \equiv \pi
r_{\text{int}}^2$, defines a typical
interaction range, $r_{\text{int}} \equiv \sqrt{\sigma/\pi}$.
On the other hand, the particle
density $n$ defines the typical interparticle distance $d \sim n^{-1/3}$.
From Eq.\ \eqref{deltanon} we conclude
that $\Delta \sim \hbar/m$, i.e., $\Delta$ is of the order the
Compton wave length of the particle. Since $\ell_{\text{mfp}} \sim (\sigma n)^{-1}$,} 
the condition we need to satisfy reads
\begin{equation} \label{condDeltaell}
\frac{\Delta}{\ell_\text{mfp}} 
\sim \frac{\Delta}{r_\text{int}}\left( \frac{r_\text{int}}{d} \right)^3 \lesssim 1\;.
\end{equation}
As long as the interparticle distance is much larger than the
interaction range, this condition is fulfilled, even if the Compton wave length
exceeds the interaction range.

{}{The dissipative currents (bulk viscous pressure, diffusion current, 
and shear-stress tensor) can be expanded in terms of powers of gradients of temperature,
chemical potential, and fluid velocity. Such a gradient expansion is,
by virtue of Eq.\ \eqref{Knudsen}, an expansion in powers of the Knudsen number.
The first-order terms in this expansion correspond to the relativistic generalization of
Navier-Stokes theory. However, it has been shown that 
this theory is acausal and unstable \cite{Pu:2009fj}. To remedy this shortcoming,
transient theories of relativistic dissipative hydrodynamics have been developed
\cite{Israel:1979wp}, where the dissipative currents relax to their Navier-Stokes values
on a time scale $\sim \ell_{\text{mfp}}$. Such theories effectively resum all orders of
the gradient expansion and render relativistic dissipative 
hydrodynamics causal and stable. However, since the values of the dissipative currents
can now differ from their asymptotic (Navier-Stokes) values, at least at early
times, besides the Knudsen number another, independent, dimensionless
quantity enters, the inverse Reynolds number $\text{R}^{-1}$, which is defined as
the ratio of a dissipative current over a quantitiy in thermodynamic equililbrium 
(e.g., pressure or particle density). Consequently, 
at early times $t \lesssim \ell_{\text{mfp}}$,
the inverse Reynolds number may differ from the Knudsen number.
However, for late times $t \gtrsim \ell_{\text{mfp}}$, all
dissipative currents relax to their Navier-Stokes values and we therefore
do not need to differentiate between Kn and $\text{R}^{-1}$.}



\section{Conclusions} 
\label{secconc}

In this paper, we provided a detailed derivation of the collision term in the 
Boltzmann equation starting from the Wigner-function formalism put forward in 
Ref.~\cite{Weickgenannt:2020aaf}. The main {}{result} of this 
work is to provide an explicit expression of the nonlocal collision kernel based on the 
framework developed in Ref.~\cite{DeGroot:1980dk}. The advantage of this formalism is 
that it relates the collision kernel to vacuum scattering amplitudes, which can be 
computed using standard field-theory techniques. Employing the 
$\hbar$, or gradient, expansion to solve for the Wigner function, it follows that the 
nonlocal term enters at next-to-leading order. Enlarging the phase space to include a 
classical variable related to spin degrees of 
freedom allows one to write the equations of motion for the Clifford components of 
the Wigner function as a single scalar Boltzmann equation. Furthermore, we proved 
that this scalar equation contains only on-shell contributions. 
{}{An implication of this work is that the conditions for the collision term to 
vanish are those of global equilibrium~\cite{Weickgenannt:2020aaf}. }
Moreover, in Refs.\ \cite{Weickgenannt:2020aaf,Speranza:2020ilk}, the relation 
between the scattering nonlocality and spin hydrodynamics is analyzed. An important 
question for phenomenological applications, which
can be addressed using nonequilibrium spin dynamics, is whether spin equilibrates 
sufficiently fast on the
time scale of the evolution of the hot and dense system created in heavy-ion collisions. 
Recent works addressing the spin-equilibration time can be found in 
Refs.~\cite{Li:2019qkf,Kapusta:2019sad,Kapusta:2019ktm,Kapusta:2020npk,Ayala:2019iin,Ayala:2020ndx}.

\section*{Acknowledgments}

The authors thank F.\ Becattini, W.\ Florkowski, X.\ Guo, U.\ Heinz,
Y.-C.\ Liu, K.\ Morawetz, R.\ Ryblewski, L.\ Tinti, G.\ Torrieri, and J.-J. Zhang 
for enlightening discussions.
The work of D.H.R., E.S., and N.W.\ is supported by the
Deutsche Forschungsgemeinschaft (DFG, German Research Foundation)
through the Collaborative Research Center CRC-TR 211 ``Strong-interaction matter
under extreme conditions'' -- project number 315477589 - TRR 211.
E.S.\ acknowledges support by
BMBF ``Forschungsprojekt: 05P2018 - Ausbau von ALICE am LHC (05P18RFCA1)".
X.-L.S. and Q.W. are supported in part by the National Natural Science Foundation 
of China (NSFC) 
under Grant No.\ 11890713 (a sub-grant of 11890710), and 11947301. X.-L.S.\ is 
supported in part by the National Natural Science Foundation of China (NSFC) 
under Grant No.\ 11890714, and 12047528.

\begin{appendix}

\section{Ensemble average of the collision term} \label{dercollapp}

Consider an arbitrary operator $O$. In kinetic theory, where dilute systems are 
considered, it is permissible to take the ensemble 
average $\langle O \rangle$ with respect to the initial, free $n$-particle states 
defined in \eq\eqref{statein}.
The ensemble average $\langle O \rangle$ is derived in Ref.~\cite{DeGroot:1980dk} 
and reads
\begin{align}
\langle O \rangle=&\sum_{n=0}^\infty \frac{4^n}{n!} \sum_{r_1 \cdots r_n}
\sum_{r^\prime_1\cdots r^\prime_n}
\int dP_1\cdots dP_n dP^\prime_1\cdots dP^\prime_n\n\\
&\times \inlanglesm \langle p_1,\ldots ,p_n;r_1,\ldots,r_n|O|  
p^\prime_1,\ldots,p^\prime_n;r^\prime_1,\ldots,r^\prime_n\rangle 
\inranglesm\, \langle a^\dagger_{\text{in} ,r_1}(p_1) \cdots a^\dagger_{\text{in} ,r_n}(p_n) 
a_{\text{in} ,r^\prime_1}(p^\prime_1)\cdots a_{\text{in} ,r^\prime_n}(p^\prime_n)\rangle \;, 
\label{opexp}
\end{align}
where $dP_i \equiv d^4p_i \,  \delta(p_i^2-m^2)$, $dP'_i\equiv d^4p_i' \, 
\delta(p_i^{\prime\,2}-m^2)$ and where we defined
\begin{align}
 &\inlanglesm \langle p_1,\ldots ,p_n;r_1,\ldots,r_n|O|  
 p^\prime_1,\ldots ,p^\prime_n;r^\prime_1,\ldots ,r^\prime_n\rangle 
 \inranglesm\n\\
&= \mathfrak{A} \sum_{m=0}^n (-1)^m \left(\frac{n !}{m! (n-m)!}\right)^2
\inlanglesm p_1, \ldots, p_m; r_1, \ldots, r_m |
p^\prime_1, \ldots, p^\prime_m; r^\prime_1, \ldots, r^\prime_n\inranglesm \n\\
&\times \inlanglesm  p_{m+1},\ldots,p_n;r_{m+1},\ldots,r_n|O|  
p^\prime_{m+1},\ldots,p^\prime_n;r^\prime_{m+1},\ldots,r^\prime_n \inranglesm\; .
\end{align}
Here the symbol $\mathfrak{A}$ indicates the antisymmetrization with respect to 
all momenta and spin indices.
Neglecting initial correlations, the expectation values of the 
creation and annihilation operators factorize according to
\begin{align}
\label{molcha}
&\langle a^\dagger_{\text{in} ,r_1}(p_1) \cdots  a^\dagger_{\text{in} ,r_n}(p_n) 
a_{\text{in} ,r^\prime_1}(p^\prime_1) \cdots a_{\text{in} ,r^\prime_n}(p^\prime_n)\rangle
=\sum_{\mathfrak{P}} (-1)^\mathfrak{P} \prod_{j=1}^n \langle a^\dagger_{\text{in} ,r_j}(p_j) 
a_{\text{in} ,r^\prime_j}(p_j^\prime)\rangle\; ,
\end{align}
where $\mathfrak{P}$ denotes the sum over all permutations of primed and 
unprimed variables with $(-1)^\mathfrak{P}=1$ for even permutations and 
$(-1)^\mathfrak{P}=-1$ for odd permutations. 

We are now interested in computing the ensemble average of the collision operator 
in Eq.~\eqref{cwww}. Using the relation for the field operator in the Heisenberg picture
\begin{equation}
\psi\left(x-\frac y2\right)= e^{\frac i\hbar P\cdot x} \psi\left(-\frac y2\right) 
e^{-\frac i\hbar P\cdot x}\; ,
\end{equation}
where $P^\mu$ is the total 4-momentum operator, and applying a similar formula 
also to $\bar{\psi},\, \rho, \bar{\rho}$, Eq.~\eqref{cwww} reads
\begin{equation}
C_{\alpha\beta} = \left\langle e^{\frac i\hbar P\cdot x} \Phi_{\alpha\beta}(p) 
e^{-\frac i\hbar P\cdot x}\right\rangle \;,\label{etfahe}
\end{equation}
with $\Phi_{\alpha\beta}$ given by Eq.~\eqref{phigen}. 
At this point we can calculate $C_{\alpha\beta}$ 
using Eq.~\eqref{opexp} with the factorization in Eq.~\eqref{molcha}. 
For the scattering kernel $C_{\alpha\beta}$, Eq.~\eqref{molcha} corresponds 
to the assumption of molecular chaos. Furthermore, since we 
consider only binary scatterings, we restrict ourselves 
to two-particle states, i.e., $n=2$. Hence, after exploiting the fact 
that two-particle states are eigenstates of the total momentum, \eq\eqref{etfahe} 
takes the form
\begin{equation}
C_{\alpha\beta}=8\sum_{r_1,r_2,r_1^\prime,r_2^\prime} 
\int dP_1 dP_2 dP_1^\prime dP_2^\prime \inlanglesm p_1,p_2;r_1,r_2| \Phi(p)| 
p_1^\prime, p_2^\prime ; r_1^\prime, r_2^\prime\inranglesm
\prod _{j=1}^2 e^{\frac i\hbar (p_j-p^\prime_j)\cdot x} 
\langle a^\dagger_{\text{in} ,r_j}(p_j)a_{\text{in} ,r^\prime_j}(p^\prime_j)\rangle \;. 
\label{kineqaad}
\end{equation}
The positive-energy part of the initial noninteracting field is
\begin{equation}
\label{psin}
\psi_\text{in}(x)=\sqrt{\frac{2}{(2\pi\hbar)^3}}\sum_{r} \int dP \,
e^{-\frac i\hbar p\cdot x} u_r(p) a_{\text{in} ,r}(p)  \;.
\end{equation}
Using the inverse relation
\begin{equation}
\frac{1}{m \sqrt{2(2 \pi \hbar)^5}} \int d^4x\, e^{\frac i\hbar p\cdot x} 
\bar{u}_r(p)\psi_\text{in}(x)
=2 \,\delta(p^2-m^2)  a_{\text{in} ,r}(p) \;, \label{replaceain}
\end{equation}
we can express Eq.~\eqref{kineqaad} in terms of the initial Wigner function
\begin{equation}
W_{\text{in},\alpha\beta}(x,p)=\int \frac{d^4y}{(2\pi\hbar)^4} e^{-\frac i\hbar p\cdot y}
  \left\langle :\bar{\psi}_{\text{in},\beta}\left(x_1\right)\psi_{\text{in},\alpha}\left(x_2\right):
  \right\rangle\; .
  \label{Wigin}
\end{equation}
The result is given in Eq.~\eqref{Wcollttt}.

\section{Calculation of the expectation value of $\Phi$} \label{collkerapp}

We want to explicitly compute the scattering-matrix element in 
Eq.~\eqref{generalcollisionterm},
\begin{equation}
\inlangle{p_1-\frac{q_1}{2},p_2-\frac{q_2}{2};r_1,r_2\Big|\Phi(p)\Big|
p_1+\frac{q_1}{2}, p_2+\frac{q_2}{2};s_1,s_2}\inrangle \;, 
\end{equation}
where the operator $\Phi(p)$ is given in \eq\eqref{phigen}.
Inserting a completeness relation of free out-states and following similar steps as done in 
Ref.~\cite{DeGroot:1980dk}, we obtain after the $y$-integration
\begin{eqnarray}
&&\inlangle{p_1-\frac{q_1}{2},p_2-\frac{q_2}{2};r_1,r_2\Big|\Phi(p)\Big|
p_1+\frac{q_1}{2}, p_2+\frac{q_2}{2};s_1,s_2}\inrangle\n\\
&=& i \sum_{r^\prime}\int dP^\prime \delta^{(4)}(p+p^\prime-p_1-p_2)\label{phi1111} \\
& \times &
\left\{\outlangle{p^\prime;r^\prime\Big|\psi(0)\Big|
p_1+\frac{q_1}{2}, p_2+\frac{q_2}{2};s_1,s_2}\inrangle 
\inlangle{p_1-\frac{q_1}{2},p_2-\frac{q_2}{2};r_1,r_2\Big|:\bar{\rho}(0):\Big|
p^\prime;r^\prime}\outrangle
\left[\gamma\cdot \left(p-\frac{q_1+q_2}{2}\right)+m\right]\right. \n \\
&  & -\left.\left[\gamma\cdot \left(p+\frac{q_1+q_2}{2}\right)+m\right] 
\outlangle{p^\prime;r^\prime\Big|:\rho(0):\Big|
p_1+\frac{q_1}{2}, p_2+\frac{q_2}{2};s_1,s_2}\inrangle 
\inlangle{p_1-\frac{q_1}{2},p_2-\frac{q_2}{2};r_1,r_2\Big|\bar{\psi}(0)\Big|
p^\prime;r^\prime}\outrangle  \right\} \;.
\n 
\end{eqnarray}
In deriving \eq\eqref{phi1111} we also made use of the fact that one- and two-particle 
states are eigenstates of the total momentum operator, hence the expectation value 
involving, e.g., $\psi(-y/2)$ is given by
\begin{eqnarray}
 \outlangle{p^\prime;r^\prime\Big|\psi\left(-\frac y2\right)\Big|
 p_1+\frac{q_1}{2}, p_2+\frac{q_2}{2};s_1,s_2}\inrangle 
 &=& e^{\frac{i}{2\hbar}(p^\prime-p_1-q_1/2 -p_2-q_2/2)\cdot y}
 \outlangle{p^\prime;r^\prime\Big|\psi(0)\Big|
 p_1+\frac{q_1}{2}, p_2+\frac{q_2}{2};s_1,s_2}\inrangle  \;. \n\\
\end{eqnarray}
In order to compute the matrix element on the right-hand side of this equation, 
we write the field $\psi(0)$ as 
a general solution of the Dirac equation in the presence of interaction 
\begin{equation}
\psi (0) = \psi_\text{in}(0) + \int d^4 x \, S_R (-x) \rho(x) \;,
\end{equation}
where $\psi_\text{in}$ is given in \eq\eqref{psin} and $S_R(x)$ is the retarded 
Green's function, which we express as a Fourier transform
\begin{equation}
S_R (x) = \frac{1}{(2\pi\hbar)^4} \int d^4p \, \tilde{S}_R(p) e^{- \frac{i}{\hbar} p \cdot x}\;,
\end{equation}
with 
\begin{equation}
 \tilde{S}_R(p)=-\frac1\hbar (\gamma\cdot p+m)G(p) \;, \label{stildeapp}
\end{equation}
and $G(p)$ defined in \eq\eqref{gapp}.
The matrix element of $\psi$ is thus given by
\begin{eqnarray}
 \outlangle{p^\prime;r^\prime\Big|\psi(0)\Big|p_1+\frac{q_1}{2}, p_2+\frac{q_2}{2};s_1,s_2}\inrangle
&= & \left[u_{s_1}\left(p_1+\frac{q_1}{2}\right)
\delta^{(3)}\left(\mathbf{p}^\prime-\mathbf{p_2}
-\frac{ \mathbf{q_2}}{2}\right)\delta_{r^\prime s_2}-(1\leftrightarrow 2)\right]
 \frac{p^{\prime0}}{\sqrt{2(2\pi\hbar)^3}}\label{phi222} \\
 & + & \tilde{S}_R\left(p_1+\frac{q_1}{2}+p_2  +\frac{q_2}{2}-p^\prime\right)
 \outlangle{p^\prime,r^\prime\Big|:\rho(0):\Big|
 p_1+\frac{q_1}{2},p_2+\frac{q_2}{2}; s_1,s_2}\inrangle \;,
\n 
\end{eqnarray}
where we made use of the orthogonality condition
$\langle p, r | p^\prime, r^\prime \rangle = p^0 \delta^{(3)}({\bf p}-{\bf p^\prime})
\delta_{r r^\prime}$. 
Plugging \eq\eqref{phi222} into \eq\eqref{phi1111} and using \eqs\eqref{stildeapp}, 
\eqref{gapp}, as well as the relation
\begin{equation}
 (\gamma \cdot k+m)_{\alpha\beta}=\sum_{r}u_r(k)_\alpha \bar{u}_r(k)_\beta\;,
\end{equation}
we obtain
\begin{eqnarray} \label{offshellscattering}
&&\inlangle{p_1-\frac{q_1}{2},p_2-\frac{q_2}{2};r_1,r_2\Big|\Phi(p)\Big|
p_1+\frac{q_1}{2}, p_2 +\frac{q_2}{2};s_1,s_2}\inrangle\n\\
&=&\frac i2\sum_{r,s} \bigg\{\frac{1}{\sqrt{2(2\pi\hbar)^3}}
\left[\delta^{(3)}\left(\mathbf{p}-\mathbf{p_1}
+\frac{\mathbf{q_2}}{2}\right)\delta\left(p^0+\sqrt{\left(\mathbf{p_2}
+\frac{\mathbf{q_2}}{2}\right)^2+m^2}-E_{p_1}-E_{p_2}\right)
\delta_{s_1r}\right.\bigg.\n\\
&&\times\left.u_{r}\left(p+\frac{q_1+q_2}{2}\right)
\bar{u}_s\left(p-\frac{q_1+q_2}{2}\right)\inlangle{p_1-\frac{q_1}{2},p_2
-\frac{q_2}{2};r_1,r_2\Big|:\bar{\rho}(0):\Big|p_2;s_2}\outrangle 
u_s\left(p-\frac{q_1+q_2}{2} \right)+(1\leftrightarrow2)\right]\bigg.\n\\
&&- \bigg.\frac{1}{\sqrt{2(2\pi\hbar)^3}}\left[\delta^{(3)}\left(\mathbf{p}-\mathbf{p_1}
-\frac{\mathbf{q_2}}{2}\right)
\delta\left(p^0+\sqrt{\left(\mathbf{p_2}-\frac{\mathbf{q_2}}{2}\right)^2+m^2}
-E_{p_1}-E_{p_2}\right)\delta_{sr_1}\right.\bigg.\n\\
&&\times\left.u_{r}\left(p+\frac{q_1+q_2}{2}\right)\bar{u}_s\left(p-\frac{q_1+q_2}{2}\right)
\bar{u}_r\left(p+\frac{q_1+q_2}{2}\right)
\outlangle{p_2;r_2\Big|:\rho(0):\Big|p_1+\frac{q_1}{2},p_2+\frac{q_2}{2};s_1,s_2}\inrangle 
+(1\leftrightarrow2)\right]\bigg.\n\\
&&\bigg.-\hbar\sum_{r^\prime}\int dP^\prime\delta^{(4)}(p+p^\prime-p_1-p_2)
\left[G\left(p+\frac{q_1+q_2}{2}\right)-G^\star\left(p-\frac{q_1+q_2}{2}\right)\right]
\bigg.\n\\
&&\times \,u_r\left(p+\frac{q_1+q_2}{2}\right)
\bar{u}_s\left(p-\frac{q_1+q_2}{2}\right)\bigg.
\bar{u}_r\left(p+\frac{q_1+q_2}{2}\right)\bigg.
\outlangle{p^\prime;r^\prime\Big|:\rho(0):\Big|p_1
+\frac{q_1}{2}, p_2+\frac{q_2}{2};s_1,s_2}\inrangle\bigg.\n\\
&&\times\bigg.\inlangle{p_1-\frac{q_1}{2},p_2-\frac{q_2}{2};r_1,r_2\Big|:\bar{\rho}(0):\Big|
p^\prime;r^\prime}
\outrangle u_s\left(p-\frac{q_1+q_2}{2}\right)\bigg\} \;,
\end{eqnarray}
where we defined $E_p\equiv \sqrt{\mathbf{p}^2+m^2}$. 
Finally, we use Eq.~\eqref{turho} to write
\begin{align} \label{offshellscatteringfinal}
 &\inlangle{p_1-\frac{q_1}{2},p_2-\frac{q_2}{2};r_1,r_2\Big|\Phi(p)\Big|p_1+\frac{q_1}{2}, p_2
 +\frac{q_2}{2};s_1,s_2}\inrangle\n\\
 &= \frac{1}{2(2\pi\hbar)^6} \sum_{r,s} u_r\left(p+\frac{q_1+q_2}{2}\right)
 \bar{u}_s\left(p-\frac{q_1+q_2}{2}\right)\, w_{r_1r_2s_1s_2}^{rs}(p_1,q_1,p_2,q_2,p)\;,
\end{align}
with
\begin{eqnarray}
 w_{r_1r_2s_1s_2}^{rs}(p_1,q_1,p_2,q_2,p)&=&2 \sum_{r^\prime}\int dP^\prime\, 
 \frac{1}{i\pi\hbar^2}\left[ G\left(p+\frac{q_1+q_2}{2}\right)
 -G^\star\left(p-\frac{q_1+q_2}{2}\right)\right]\n\\
 &&\times\delta^{(4)}(p+p^\prime-p_1-p_1)\Big\langle{p+\frac{q_1+q_2}{2}, 
 p^\prime;r,r^\prime\Big|t\Big|p_1+\frac{q_1}{2}, p_2+\frac{q_2}{2};s_1,s_2}\Big\rangle\n\\
 &&\times\Big\langle{p_1-\frac{q_1}{2},p_2-\frac{q_2}{2};r_1,r_2\Big|t^\dagger\Big|
 p-\frac{q_1+q_2}{2},p^\prime;s,r^\prime}\Big\rangle \n\\
  &&-i2\pi\hbar \delta^{(3)}\left(\mathbf{p}-\mathbf{p_1}+\frac{\mathbf{q_2}}{2}\right)
  \delta\left(p^0+p_2^0+\frac{q_2^0}{2}-E_{p_1}-E_{p_2}\right)\n\\
 &&\times\Big\langle{p_1-\frac{q_1}{2},p_2-\frac{q_2}{2};r_1,r_2\Big|t^\dagger\Big|
 p-\frac{q_1+q_2}{2},p_2;s,s_2}\Big\rangle\delta_{rs_1}+(1\leftrightarrow2)\n\\
 &&+i2\pi\hbar \delta^{(3)}\left(\mathbf{p}-\mathbf{p_1}-\frac{\mathbf{q_2}}{2}\right)
 \delta\left(p^0+p_2^0-\frac{q_2^0}{2}-E_{p_1}-E_{p_2}\right)\n\\
 &&\times\Big\langle{p_1+\frac{q_1}{2},p_2+\frac{q_2}{2};r,r_2\Big|t\Big|
 p+\frac{q_1+q_2}{2},p_2;s_1,s_2}\Big\rangle \delta_{r_1s}+(1\leftrightarrow2)\; . 
 \label{w1app}
\end{eqnarray}
Here we made use of the fact that to linear order in $\mathbf{q}_2$ we may replace 
$\sqrt{(\mathbf{p_2}\pm \mathbf{q_2}/2)^2+m^2}= p_2^0\pm q_2^0/2$. Working to linear order is
sufficient since we only consider zeroth and first order terms of a Taylor expansion in $q_i$ in 
Eq.\ (\ref{generalcollisionterm}). In this form, the scattering-matrix element is inserted into 
Eq.~\eqref{generalcollisionterm}. 
Furthermore, for the sake of simplicity  we define  
\begin{equation}
\label{b13}
 w_{r_1r_2s_1s_2}^{rs}(p_1,p_2,p)\equiv w_{r_1r_2s_1s_2}^{rs}(p_1,q_1=0,p_2,q_2=0,p)\;.
\end{equation}

\section{Spinor identities}

For any on-shell momentum $p^\mu = (E_p, \mathbf{p})$, we can write
\begin{equation}
 u_r(p)=\frac{\gamma\cdot p+m}{\sqrt{2m(E_p+m)}}\, u_r(p_\star)\;,
\end{equation}
where $p_\star^\mu = (m,\mathbf{0})$ is the 4-momentum in the rest frame of the particle.
Then, the following identities hold to first order in $q^\mu$,
\begin{align} \label{uu}
 \bar{u}_s\left(p+\frac{q}{2}\right)u_r\left(p-\frac{q}{2}\right) & = 2m\, \delta_{sr}-\frac{i}{2(E_p+m)}\, q_\mu p_\nu \Sigma^{\mu\nu}_{sr}(p_\star) \;,\n\\
  \bar{u}_s\left(p+\frac{q}{2}\right)\gamma^\alpha u_r\left(p-\frac{q}{2}\right)
 &= 2p^\alpha\delta_{sr}+\frac{im}{2(E_p+m)}\, q_\mu \Sigma^{\alpha\mu}_{sr}(p_\star)
 -\frac{i}{E_p+m}\epsilon^{\alpha\mu\nu\rho}q _\mu p_\nu n_{sr \rho}(p_\star) \;,\n\\
 \bar{u}_s\left(p+\frac{q}{2}\right)\gamma^5\gamma^\alpha u_r\left(p-\frac{q}{2}\right)
 &= 2m n^\alpha_{sr}(p)-\frac{i}{E_p+m}\epsilon^{\alpha\mu\nu0}q_\nu p_\mu \delta_{sr} \;,
\end{align}
where we used Eqs.\ (\ref{n_mu_rs}), (\ref{Sigma_munnu_rs}), and the identity
\begin{equation}
 \gamma^\mu\gamma^\alpha\gamma^\nu
 =g^{\mu\alpha}\gamma^\nu+g^{\alpha\nu}\gamma^\mu
 -g^{\nu\mu}\gamma^\alpha-i\epsilon^{\mu\alpha\nu\rho}\gamma_\rho\gamma^5\;.
\end{equation}

\section{Calculation of nonlocal collision term} \label{non_loc_coll_calc}

The second contribution of the nonlocal term in \eq\eqref{cnlsecond} term is given by 
\begin{align} \label{secondcontributiontofirstordercollision}
m\, \mathfrak{C}^{(1)}_{nl,2}={}&i\frac{(2\pi\hbar)^6}{8m^4}\sum_{r_1,r_2,s_1,s_2}
\int d^4p_1 d^4p_2d^4q_1 d^4q_2 \, \delta^{(4)}(q_1)\delta^{(4)}(q_2) \n\\
  &\times\bigg\{ \partial_{q_1}^\mu
  \Tr\left[\left(\frac{m}{p^2}\, p\cdot\gamma-\ms \cdot \gamma \, \gamma^5\right)
  \inlangle{p_1-\frac{q_1}{2},p_2-\frac{q_2}{2};r_1,r_2\Big|\Phi(p)\Big|
  p_1+\frac{q_1}{2}, p_2+\frac{q_2}{2};s_1,s_2}\inrangle\right]\bigg.\n\\
  & \times\bar{u}_{s_2}(p_2)W(x,p_2)u_{r_2}(p_2)\bar{u}_{s_1}(p_1)
  \partial_{\mu}W(x,p_1)u_{r_1}(p_1)\n\\
  &+\partial_{q_2}^\mu
  \Tr\left[\left(\frac{m}{p^2}\, p\cdot\gamma-\ms \cdot \gamma\, \gamma^5\right)
  \inlangle{p_1-\frac{q_1}{2},p_2-\frac{q_2}{2};r_1,r_2\Big|\Phi(p)\Big|
  p_1+\frac{q_1}{2}, p_2+\frac{q_2}{2};s_1,s_2}\inrangle\right]\n\\
  & \times\bigg.\bar{u}_{s_1}(p_1)W(x,p_1)u_{r_1}(p_1)\bar{u}_{s_2}(p_2)
  \partial_{\mu}W(x,p_2)  u_{r_2}(p_2)\bigg\}\n\\
  ={}& \frac{i}{16m^4} \sum_{r,s,r_1,r_2,s_1,s_2}\int d^4p_1 d^4p_2d^4q_1 d^4q_2 \,
  \delta^{(4)}(q_1)\delta^{(4)}(q_2) \n\\
  &\times\Bigg\{\partial_{q_1}^\mu\Bigg[\bar{u}_s\left(p-\frac{q_1}{2}-\frac{q_2}{2}\right)
  \left(\frac{m}{p^2}\,p\cdot\gamma-\ms \cdot \gamma\,\gamma^5\right)
  u_r\left(p+\frac{q_1+q_2}{2}\right) w_{r_1r_2s_1s_2}^{rs}(p_1,q_1,p_2,q_2,p)\Bigg.\Bigg]
  \n\\
 &\times \bar{u}_{s_2}(p_2)W(x,p_2)u_{r_2}(p_2)\bar{u}_{s_1}(p_1)
 \partial_{\mu}W(x,p_1)u_{r_1}(p_1)\n\\
 &+\partial_{q_2}^\mu\Bigg[\bar{u}_s\left(p-\frac{q_1+q_2}{2}\right)
 \left(\frac{m}{p^2}\, p\cdot\gamma-\ms \cdot \gamma\,\gamma^5\right)
 u_r\left(p+\frac{q_1}{2}+\frac{q_2}{2}\right)w_{r_1r_2s_1s_2}^{rs}(p_1,q_1,p_2,q_2,p)
 \Bigg.\Bigg]\n\\
 &\times\Bigg.\bar{u}_{s_1}(p_1)W(x,p_1)u_{r_1}(p_1)\bar{u}_{s_2}(p_2)
 \partial_{\mu}W(x,p_2)  u_{r_2}(p_2)\Bigg\}\n\\
 ={}& \frac{i}{16m^4} \sum_{r,s,r_1,r_2,s_1,s_2}\int d^4p_1 d^4p_2 
 \Bigg\{\frac{i}{2(p^0+m)}\left[p_\nu\Sigma_{sr}^{\mu\nu}(p_\star)+
 \epsilon^{\nu\lambda\mu0}p_\nu\ms_\lambda\delta_{sr}\right]
 w_{r_1r_2s_1s_2}^{rs}(p_1,p_2,p)\n\\
 &\times\Bigg.\partial_{\mu}\bar{u}_{s_1}(p_1)W(x,p_1)u_{r_1}(p_1)
 \bar{u}_{s_2}(p_2)W(x,p_2)  u_{r_2}(p_2)\Bigg\}\n\\
  &+\frac{i}{16 m^4}\sum_{r,s,r_1,r_2,s_1,s_2}\int d^4p_1 d^4p_2
  \left[\frac{m}{p^2}\bar{u}_{s}(p)\, p\cdot\gamma\, u_{r}(p)
  -\ms_\mu\bar{u}_{s}(p)\gamma^\mu\gamma^5 u_{r}(p)\right]\n\\
  &\times\left\{ \left[\partial_{q_1}^\mu 
  w_{r_1 r_2 s_1 s_2}^{rs}(p_1,q_1,p_2,q_2,p)  \right]_{q_1=q_2=0}
  \bar{u}_{s_2}(p_2)W(x,p_2)u_{r_2}(p_2)\bar{u}_{s_1}(p_1)\partial_{\mu}W(x,p_1)
  u_{r_1}(p_1) \right. \n\\
  &+\left. \left[\partial_{q_2}^\mu 
  w_{r_1 r_2 s_1 s_2}^{rs}(p_1,q_1,p_2,q_2,p)\right]_{q_1=q_2=0}
  \bar{u}_{s_1}(p_1)W(x,p_1)u_{r_1}(p_1)\bar{u}_{s_2}(p_2)
  \partial_{\mu}W(x,p_2)u_{r_2}(p_2)\right\}\;,
\end{align}
where we used \eq\eqref{offshellscatteringfinal} in the second step and, in the last step, 
\eq\eqref{b13} and the relation 
\begin{align}
 &\partial_{q_{j}}^\alpha\left[\frac{m}{p^2}p_\mu
 \bar{u}_s\left(p-\frac{q_1+q_2}{2}\right)\gamma^\mu u_r\left(p+\frac{q_1+q_2}{2}\right)
 +\ms_\mu\bar{u}_s\left(p-\frac{q_1+q_2}{2}\right)\gamma^5\gamma^\mu 
 u_r\left(p+\frac{q_1+q_2}{2}\right)\right]_{q_1=q_2=0}\n\\
 &=\frac{i}{2(p^0+m)}\left[ \frac{m^2}{p^2} p_\nu\Sigma_{sr}^{\alpha\nu}(p_\star)
 +\epsilon^{\nu\mu\alpha0}p_\nu\ms_\mu \delta_{sr} \right]\;.
\end{align}
When inserting this equation into Eq.\ (\ref{secondcontributiontofirstordercollision}), 
the term $m^2/p^2 = 1$, since 
$w_{r_1r_2s_1s_2}^{rs}(p_1,p_2,p)$ puts the 4-momentum $p^\mu$ on-shell, 
see \eq\eqref{wwww}.

The $q_j$-derivatives acting on $w_{r_1 r_2 s_1 s_2}^{rs}(p_1,q_1,p_2,q_2,p)$ 
in the last two lines of Eq.\ \eqref{secondcontributiontofirstordercollision} contain 
several terms. In order to calculate them, 
it is convenient to split $w_{r_1 r_2 s_1 s_2}^{rs}(p_1,q_1,p_2,q_2,p)$, 
cf.\ \eq\eqref{w1app}, into a gain term,
\begin{eqnarray}
\lefteqn{\hspace*{-0.4cm} w_{r_1r_2s_1s_2,\text{gain}}^{rs}(p_1,q_1,p_2,q_2,p)=
2\sum_{r^\prime}\int dP^\prime\, 
\frac{1}{i\pi\hbar^2}\left[ G\left(p+\frac{q_1+q_2}{2}\right)
-G^\star\left(p-\frac{q_1+q_2}{2}\right)\right]\delta^{(4)}(p+p^\prime-p_1-p_1)}\n\\
 &\times&
 \Big\langle{p+\frac{q_1+q_2}{2},p^\prime;r,r^\prime\Big|t\Big|
 p_1+\frac{q_1}{2}, p_2+\frac{q_2}{2};s_1,s_2}\Big\rangle
 \Big\langle{p_1-\frac{q_1}{2},p_2-\frac{q_2}{2};r_1,r_2\Big|t^\dagger\Big|
 p-\frac{q_1+q_2}{2},p^\prime;s,r^\prime}\Big\rangle \;, \label{gain}
\end{eqnarray}
and a loss term
\begin{eqnarray}
w_{r_1r_2s_1s_2,\text{loss}}^{rs}(p_1,q_1,p_2,q_2,p)&=& -i2\pi\hbar \delta^{(3)}
\left(\mathbf{p}-\mathbf{p_1}+\frac{\mathbf{q_2}}{2}\right)\delta\left(p^0
+\sqrt{\left(\mathbf{p_2}+\frac{\mathbf{q_2}}{2}\right)^2+m^2}-E_{p_1}-E_{p_2}\right)\n\\
 &&\times\Big\langle{p_1-\frac{q_1}{2},p_2-\frac{q_2}{2};r_1,r_2\Big|t^\dagger\Big|
 p-\frac{q_1+q_2}{2},p_2;s,s_2}\Big\rangle\delta_{rs_1}+(1\leftrightarrow2)\n\\
 &&+i2\pi\hbar \delta^{(3)}\left(\mathbf{p}-\mathbf{p_1}-\frac{\mathbf{q_2}}{2}\right)
 \delta\left(p^0+\sqrt{\left(\mathbf{p_2}-\frac{\mathbf{q_2}}{2}\right)^2+m^2}
 -E_{p_1}-E_{p_2}\right)\n\\
 &&\times\Big\langle{p_1+\frac{q_1}{2},p_2+\frac{q_2}{2};r,r_2\Big|t\Big|
 p+\frac{q_1+q_2}{2},p_2;s_1,s_2}\Big\rangle\delta_{r_1s}+(1\leftrightarrow2)\;. \label{loss}
\end{eqnarray}
Since we compute a contribution of order $\mathcal{O}(\hbar)$, the
Wigner functions in Eq.\ (\ref{secondcontributiontofirstordercollision}) 
can be approximated by their
zeroth-order expression, such that the terms $\sim W \partial_\mu W$ 
will give rise to terms
$\sim f^{(0)} \partial_\mu f^{(0)}$, with $f$ being the zeroth-order contribution to
$f(x,p,\ms)$. To zeroth order, the $\ms$-dependence vanishes, such that 
$f^{(0)}(x,p,\ms) \equiv f^{(0)}(x,p)$.
Acting with the $q_j$-derivative on the gain part, Eq.\ (\ref{gain}), the respective terms
in Eq.\  (\ref{secondcontributiontofirstordercollision}) lead to contributions of the form
\begin{align}
 &\left[f^{(0)}(x,p_2)\partial^\mu f^{(0)}(x,p_1)\partial_{q_1\mu}
 +f^{(0)}(x,p_1)\partial^\mu f^{(0)}(x,p_2)\partial_{q_2\mu}\right] 
  \times \left[ G\left(p+\frac{q_1+q_2}{2}\right)
-G^\star\left(p-\frac{q_1+q_2}{2}\right)\right]_{q_1=q_2=0}\n\\
 &= -\frac12\hbar^2\partial^\mu f^{(0)}(x,p_1)f^{(0)}(x,p_2) 
 \partial_{q\mu}\left[\frac{1}{(p+q)^2-m^2-i\epsilon(p^0+q^0)}
 -\frac{1}{(p-q)^2-m^2+i\epsilon(p^0-q^0)}\right]_{q=0}\n\\
&=\partial^\mu f^{(0)}(x,p_1)f^{(0)}(x,p_2)\frac{p_\mu}{p^2-m^2}
\left[ G(p) + G^\star(p)\right]
 \;. \label{firstoffshellcontributionfrominteractions}
\end{align}
Due to the factor $p^2-m^2$ in the denominator, this is an off-shell contribution 
to the Boltzmann equation.
Further off-shell contributions also emerge when the $q_j$-derivatives act on the 
loss term, Eq.\ (\ref{loss}), i.e., in terms of the form
\begin{eqnarray}
 &&\frac{im}{2}\sum_{r,s,r_1,r_2,s_1,s_2}\int dP_1 dP_2\, h_{sr}(p,\ms_1) 
 \left[f^{(0)}(x,p_2)\partial_{\nu}f^{(0)}(x,p)\partial_{q_1}^\nu
 +f^{(0)}(x,p)\partial_{\nu}f^{(0)}(x,p_2)\partial_{q_2}^\nu\right]\n\\
 &&\times 
 \Big[-i2\pi\hbar\delta^{(3)}\left(\mathbf{p}-\mathbf{p_1}+\frac{\mathbf{q_2}}{2}\right)
 \delta\left(p^0+p_2^0+\frac{q_2^0}{2}-E_{p_1}-E_{p_2}\right)\Big.\n\\
 &&\times\Big\langle{p_1-\frac{q_1}{2},p_2-\frac{q_2}{2};r_1,r_2\Big|t^\dagger\Big|
 p-\frac{q_1+q_2}{2},p_2-\frac{q_2}{2};s,s_2}\Big\rangle\delta_{rs_1}
 +(1\leftrightarrow2)\n\\
 &&+i2\pi\hbar \delta^{(3)}\left(\mathbf{p}-\mathbf{p_1}-\frac{\mathbf{q_2}}{2}\right)
 \delta\left(p^0+p_2^0-\frac{q_2^0}{2}-E_{p_1}-E_{p_2}\right)\n\\
 &&\Big.\times\Big\langle{p_1+\frac{q_1}{2},p_2+\frac{q_2}{2};r,r_2\Big|t\Big|
 p+\frac{q_1+q_2}{2},p_2+\frac{q_2}{2};s_1,s_2}
 \Big\rangle\delta_{r_1s}
 +(1\leftrightarrow2)\Big]_{q_1=q_2=0}\delta_{r_1s_1}\delta_{r_2s_2}\n\\
 &=& \frac{im}{2}\sum_{r,s,r_1,r_2,s_2}\int dP_{2} h_{sr}(p,\ms_1) 
 {\partial_{\nu} \left[f^{(0)}(x,p_2)f^{(0)}(x,p)\right]}
 \left(\partial_{q_1}^\nu+\partial_{q_2}^\nu\right)\n\\
 &&\times \Big[- {\frac{i\pi\hbar}{E_{p+\frac{q_2}{2}}}} \delta\left(p^0+\frac{q_2^0}{2}
 -E_{p+\frac{q_2}{2}}\right)
 \Big\langle{p-\frac{q_1-q_2}{2},p_2-\frac{q_2}{2};r_1,r_2\Big|t^\dagger\Big|
 p-\frac{q_1+q_2}{2},p_2-\frac{q_2}{2};s,s_2}\Big\rangle 
 \delta_{rr_1}\delta_{r_2s_2}\n\\
 &&+{\frac{i\pi\hbar}{E_{p-\frac{q_2}{2}}}} 
 \delta\left(p^0-\frac{q_2^0}{2}-E_{p-\frac{q_2}{2}}\right)
 \Big\langle{p+\frac{q_1-q_2}{2},p_2+\frac{q_2}{2};r,r_2\Big|t\Big|
 p+\frac{q_1+q_2}{2},p_2+\frac{q_2}{2};r_1,s_2}
 \Big\rangle\delta_{r_1s}\delta_{r_2s_2}\Big]_{q_1=q_2=0} \;.\n\\ 
\end{eqnarray} 
It is clear that both on- and off-shell contributions are present, since
\begin{align}
\partial_q^\mu \left. {\frac{1}{2E_{p+\frac{q}{2}}}} \delta\left(p^0+\frac{q^0}{2}-E_{p+\frac{q}{2}}\right) \right|_{q=0}={}& \partial_q^\mu \left. \delta \left(\left(p^0+\frac{q^0}{2}\right)^2-E^2_{p+\frac{q}{2}}\right) \right|_{q=0} \n \\
&{}=p^\mu \delta^\prime (p^2-m^2) \;.
\end{align}
We can collect all the off-shell contributions to the collision term as
\begin{eqnarray}
\mC_{\text{off-shell}}^{(1)}& =& \frac{i}{2(p^2-m^2)}\, p \cdot \partial \sum_{r,s,r_1,r_2,s_1,s_2} dP_2 \, 
h_{sr}(p,\ms) f^{(0)}(x,p_1)  f^{(0)}(x,p_2) \n\\
& \times & \bigg\{ 2\sum_{r^\prime}\int dP_1 dP^\prime\, \frac{1}{i\pi\hbar^2}\left[G(p)+G^\star(p)\right]
\delta^{(4)}(p+p^\prime-p_1-p_1)  \n \\
&& \times \langle{p,p^\prime;r,r^\prime|t|p_1, p_2;s_1,s_2}\rangle\langle{p_1,p_2;r_1,r_2|t^\dagger|p,p^\prime;s,r^\prime}\rangle \n\\
&&  + i2\pi\hbar\, p^0 \delta(p^2-m^2) \left[\langle{p,p_2;r_1,r_2|t^\dagger|p,p_2;s,s_2}\rangle\delta_{rr_1}\delta_{s_2r_2}+\langle{p,p_2;r,r_2|t|p,p_2;r_1,s_2}\rangle\delta_{r_1s}\delta_{r_2s_2}\right] \bigg\} \;. \label{c2offshell}
\end{eqnarray}

We now show that the off-shell part  \eq\eqref{c2offshell} cancels 
with the off-shell part on the left-hand side 
of the Boltzmann equation \eqref{Boltzmannnn}. Using the quasiparticle approximation in 
\eq\eqref{onshellsolution}, the left-hand side of \eq\eqref{Boltzmannnn} is given by
\begin{equation}
m\, p\cdot\partial\, \delta(p^2-m^2-\hbar\delta m^2)f(x,p,\ms)
={}m \delta(p^2-m^2) p\cdot \partial \, f(x,p,\ms) 
+\hbar\frac{1}{p^2-m^2}\, p\cdot \partial \, \mathfrak{M}^{(0)} \;, 
  \end{equation}
where the correction to the mass shell at zeroth order reads
\begin{align}
 \mathfrak{M}^{(0)}=
 {}&  \frac{im}{2} \sum_{r,s,r_1,r_2,s_1,s_2}\int dP_1\, dP_2\, h_{sr}(p,\ms)\,
 \mathfrak{m}_{r_1,r_2,s_1,s_2}^{rs}(p_1,p_2,p)\prod_{j=1}^2 \delta_{s_jr_j}f^{(0)}(x,p_j) \;,
    \label{deltaMsq0}
\end{align}
with 
\begin{eqnarray}
 \mathfrak{m}_{r_1,r_2,s_1,s_2}^{r,s}(p_1,p_2,p)&=&
2 \sum_{r^\prime}\int dP^\prime\, \frac{1}{i\pi\hbar^2}
\left[ G\left(p\right)+G^\star\left(p\right)\right]
\delta^{(4)}(p+p^\prime-p_1-p_1)\n\\
 &&\times\langle{p,p^\prime;r,r^\prime|t|p_1, p_2;s_1,s_2}\rangle
 \langle{p_1,p_2;r_1,r_2|t^\dagger|p,p^\prime;s,r^\prime}\rangle\n \\
& & +i2\pi\hbar p^0\delta(p^2-m^2) \{\delta^{(3)}(\mathbf{p}-\mathbf{p_1})
 [\langle{p_1,p_2;r_1,r_2|t^\dagger|p,p_2;s,s_2}\rangle\delta_{rs_1} \n \\
&  & +\langle{p_1,p_2;r,r_2|t|p,p_2;s_1,s_2}\rangle\delta_{r_1s}]+(1\leftrightarrow 2)\}\; .
\end{eqnarray}
The steps to obtain \eq\eqref{deltaMsq0} are completely analogous to the 
calculation that leads to the local collision term, since we see from Eqs.~\eqref{deltaM} 
and \eqref{wigboltz9} that $\delta M$ is just the real part of the quantity of which 
$C$ given by Eq.~\eqref{generalcollisionterm} is the imaginary part.

Comparing \eq\eqref{c2offshell} with \eq\eqref{deltaMsq0} we find up to first order
\begin{equation}
m\, \mC_{\text{off-shell}}^{(1)}=\frac{1}{p^2-m^2} \, p\cdot \partial \, \mathfrak{M}^{(0)}\; ,
\end{equation}
which implies that all off-shell contributions cancel on the left- and right-hand sides and
the Boltzmann equation involves only on-shell terms.
Thus, we obtain the following kinetic equation for the distribution function $f(x,p,\ms)$ 
\begin{equation}
 \delta(p^2-m^2)\, p\cdot\partial \, f(x,p,\ms)
 = \delta(p^2-m^2)\mathfrak{C}_{\text{on-shell}}[f] \;,
\end{equation}
with
\begin{equation}
 \mC_{\text{on-shell}}[f]\equiv \mC_{\osl}[f]+\hbar\, \mC_{\os,nl,1}^{(1)}[f]
 +\hbar\, \mC_{\os,2,1}^{(1)}[f]+\hbar\, \mC^{(1)}_{\os,2,2}[f] \;.
\end{equation}

Here,
 we obtained from the first two lines in the last equality in 
 \eq\eqref{secondcontributiontofirstordercollision}
\begin{eqnarray}
 \mathfrak{C}^{(1)}_{\os,2,1}[f]&=& -\frac{1}{8m(p^0+m)}  \sum_{r,s,r^\prime,r_1,r_2} 
  \left(p_\nu\Sigma_{sr}^{\mu\nu}(p_\star)+\epsilon^{\nu\lambda\mu0}
  p_\nu\ms_\lambda\delta_{sr}\right)\n\\
&& \times \int dP_1\, dP_2\, dP^\prime\,   \delta^{(4)}(p+p^\prime-p_1-p_2) 
  \langle{p,p^\prime;r,r^\prime|t|p_1,p_2;r_1,r_2}\rangle 
  \langle{p_1,p_2;r_1,r_2|t^\dagger|p,p^\prime;s,r^\prime}\rangle \n\\
  &&\times\left[ \partial_{\mu} f^{(0)}(x,p_1)f^{(0)}(x,p_2)
  -\partial_{\mu} f^{(0)}(x,p^\prime)f^{(0)}(x,p)  \right] \;, 
  \label{collvanish}
  \end{eqnarray}
where we properly relabeled indices and applied the optical theorem \eq\eqref{opttheo}. 
Furthermore, the on-shell contribution from the 
last three lines in \eq\eqref{secondcontributiontofirstordercollision} is given by
\begin{eqnarray}
 \mC_{\os,2,2}^{(1)}[f]
 &=&\frac{1}{4m}\sum_{r_1,r_2,s_1,s_2}\sum_{r,r^\prime,s}\int dP_1\, dP_2\, dP^\prime\,
 h_{sr}(p,\ms)\delta^{(4)}(p+p^\prime-p_1-p_2)\delta(p^2-m^2)\n\\
 &&\times \left[f(x,p_2)\partial_{\nu}f(x,p_1)\partial_{q_1}^\nu
 +f(x,p_1)\partial_{\nu}f(x,p_2)\partial_{q_2}^\nu\right]
 \Big\langle{{p+\frac{q_1}{2}+\frac{q_2}{2}},{p^\prime};r,r^\prime\Big|t\Big|
 {p_1+\frac{q_1}{2}}, {p_2+\frac{q_2}{2}};s_1,s_2}\Big\rangle\n\\
 &&\times\Big\langle{{p_1-\frac{q_1}{2}},{p_2-\frac{q_2}{2}};r_1,r_2\Big|t^\dagger\Big|
 {p-\frac{q_1}{2}-\frac{q_2}{2}},{p^\prime};s,r^\prime}\Big\rangle 
 \delta_{s_1r_1}\delta_{s_2r_2}\n\\
 &&-\frac{m}{16\pi}\sum_{r_2,s_2}\int dP_{2}\sum_{r,s} h_{sr}(p,\ms)\delta_{r_2s_2} 
 \delta(p^2-m^2){\partial_{\nu} \left[f(x,p_2)f(x,p)\right]}
 \left(\partial_{q_1}^\nu+\partial_{q_2}^\nu\right)\n\\
 &&\times i4\pi\hbar  \Big\langle{p+\frac{q_2}{2}-\frac{q_1}{2},p_2-\frac{q_2}{2};r,r_2\Big|
 t+t^\dagger\Big| p-\frac{q_1}{2}-\frac{q_2}{2},p_2-\frac{q_2}{2};s,s_2}\Big\rangle \;.
 \label{cmomder}
\end{eqnarray}
As discussed in 
Sec.~\ref{secnon}, in accordance with the low-density approximation 
we neglect the momentum derivatives of the scattering amplitude and, hence, all terms in 
\eq\eqref{cmomder} vanish. We now show that the term in \eq\eqref{cmomder} vanishes 
once the zeroth-order distribution function is inserted.
The zeroth-order distribution function make the zeroth-order collision term 
$\mC^{(0)}$ vanish and is given by the usual Boltzmann form
\begin{equation}
\label{zerof}
 f^{(0)}(x,p)=\frac{1}{(2\pi\hbar)^3}e^{-\beta(x)\cdot p}\;.
\end{equation}
(We consider here the simplest case of a neutral fluid. 
Adding a chemical potential is trivial and does not change the conclusion.) 
Inserting \eq\eqref{zerof} into Eq.~\eqref{collvanish}, we find
\begin{widetext}
 \begin{eqnarray}
    \mathfrak{C}^{(1)}_{\os,2,1}[f]&=&\frac{1}{(2\pi\hbar)^3}\frac{1}{8m(p^0+m)}  
    \sum_{r,s,r^\prime,r_1,r_2}  \left[p_\nu\Sigma_{sr}^{\mu\nu}(p_\star)+
    \epsilon^{\nu\lambda\mu0}p_\nu\ms_\lambda\delta_{sr}\right]\n\\
&& \times \int dP_1\, dP_2\, dP^\prime\,   \delta^{(4)}(p+p^\prime-p_1-p_2)   
  \langle{p,p^\prime;r,r^\prime|t|p_1,p_2;r_1,r_2}\rangle 
  \langle{p_1,p_2;r_1,r_2|t^\dagger|p,p^\prime;s,r^\prime}\rangle \n\\
  &&\times\partial_\mu\beta_\lambda(p_1^\lambda+p_2^\lambda-p^\lambda 
  -p^{\prime\lambda})e^{-\beta\cdot (p_1+p_2)}\n\\
  &=&0\;.
  \label{ceq0}
\end{eqnarray}
\end{widetext}
Therefore, we proved the structure of the Boltzmann equation and the on-shell collision 
terms given in \eqs\eqref{onshellcolll} and \eqref{coll123}.
\end{appendix}

\bibliography{biblio_paper_long}{}

\end{document}